\documentclass[12pt]{iopart}
\usepackage{iopams}
\usepackage{graphicx}
\usepackage{amssymb}

\usepackage{color}

\begin{document}
                 
\title{Decoherence times of universal two-qubit gates in the presence of broad-band noise} 
\author{E. Paladino,}
\address{Dipartimento di Fisica e Astronomia, Universit\`a di Catania
and CNR IMM MATIS, Catania, 
C/O Viale Andrea Doria 6, Ed.10, 95125 Catania, Italy.}
\ead{epaladino@dmfci.unict.it}
\author{A. D'Arrigo,}
\address{Dipartimento di Fisica e Astronomia, Universit\`a di Catania
and CNR IMM MATIS, Catania, 
C/O Viale Andrea Doria 6, Ed.10, 95125 Catania, Italy.}
\author{A. Mastellone,}
\address{C.I.R.A. Centro Italiano Ricerche Aerospaziali -
Via Maiorise snc - 81043 Capua, Caserta, Italy.}
\author{G. Falci}
\address{Dipartimento di Fisica e Astronomia, Universit\`a di Catania
and CNR IMM MATIS, Catania, 
C/O Viale Andrea Doria 6, Ed.10, 95125 Catania, Italy.}

\begin{abstract}
Controlled generation of entangled states of two quantum bits 
is a fundamental step toward the implementation of a quantum 
information processor. In nano-devices this operation is
counteracted by the solid-state environment,
characterized by broadband and non-monotonic power spectrum 
often $1/f$ at low frequencies.
For single qubit gates, incoherent processes due to fluctuations 
acting on different time scales result in peculiar short- and 
long-time behaviors.
Markovian noise originates exponential decay with relaxation
and decoherence times, $T_1$ and $T_2$, simply related to the 
symmetry of the qubit-environment coupling Hamiltonian. Noise 
with $1/f$ power spectrum at low frequencies is instead responsible for
defocusing processes and algebraic short-times behavior.
In this article we identify the relevant decoherence times of an 
entangling operation due to the different decoherence channels 
originated from solid state noise.
Entanglement is quantified by the concurrence, which we evaluate in analytic form 
employing a multi-stage approach. "Optimal" operating conditions of reduced sensitivity
to noise sources are identified. We apply this analysis to a superconducting 
$\sqrt{{\rm i-SWAP}}$ gate for experimental noise spectra.
\end{abstract}

%\keywords{decoherence; 1/f-noise;}
\submitto{\NJP}

\maketitle

\section{Introduction}

The implementation of a universal two-qubit gate involving an entanglement
operation on two quantum bits represents a necessary step toward the
construction of a scalable quantum computer~\cite{Nielsen}.   
Intense research on solid state nano-devices during the last decade
has established the possibility to combine quantum coherent behavior with 
the existing integrated-circuit fabrication technology.     
In particular, based  on superconducting technologies, a variety of 
high-fidelity single qubit gates are nowadays available~\cite{single-super,vion,Clarke08}, 
two-qubit logic gates~\cite{coupled-exp-fix,coupled-exp} and violations of Bell's 
inequalities~\cite{Bell-violation} have been demonstrated, 
high-fidelity Bell states generated~\cite{Bell-generation}.
The recent demonstrations of simple quantum algorithms~\cite{DiCarlo09}
and  three-qubit entanglement~\cite{three}  are further important
steps toward a practical quantum computation with superconducting circuits.

The requirements for building an elementary quantum processor are however
quite demanding on the efficiency of the protocols. This includes both a severe 
constraint on readout and a sufficient isolation from fluctuations to reduce 
decoherence effects. 
Solid-state noise sources are often characterized by broad-band and 
non-monotonic power spectrum. 
Similar noise characteristics have been reported in implementations
based on Cooper-pair-boxes (CPB)~\cite{nak-echo,nak-spectrum,ithier}, 
in persistent current~\cite{Bylander11} and phase qubits~\cite{Simmonds,Ustinov}.
Usually, the spectrum of at least one of the noise sources is $1/f$ at 
low-frequencies~\cite{nak-spectrum,Bylander11,Kafanov,Koch,VanHarlingen,Bialczak}. 
At the system's eigen-frequencies instead ($5-15$~GHz) 
indirect measurements indicate 
white or ohmic spectrum~\cite{nak-spectrum,ithier,Bylander11}. 
Sometimes spurious resonances of various physical origin have  been 
observed~\cite{Simmonds,Ustinov,resonators}.
 
At the single-qubit level, the effects of the environmental degrees of freedom 
responsible for the various parts of the spectrum have been
clearly identified leading to a convenient classification in terms of  
{\em quantum noise} and {\em adiabatic noise} effects~\cite{ithier,PRL05}.
Understanding how these mechanisms affect an entanglement-generating two-qubit
gate is a relevant issue not yet investigated and it is the subject of the present
article.

The picture for a single qubit can be summarised as follows.
Noise at frequencies of the order of the system's splittings may induce 
incoherent energy exchanges between qubit and environment ({\em quantum noise}).
Relaxation processes occur only if the qubit-environment interaction induces
spin flips in the qubit eigenbasis, i.e. for transverse noise.
Weakly-coupled Markovian noise can be treated by a Born-Markov master equation~\cite{weiss}.
It leads to relaxation and decoherence times denoted  respectively $T_1$ and $T_2$ in Nuclear 
Magnetic  Resonance (NMR)~\cite{shlichter}.
For transverse noise they are related by $T_2=2T_1$. Longitudinal noise does not
induce spin flips, but it is responsible for pure dephasing  with 
a decay-time denoted $T_2^*$~\cite{shlichter}. 
In general, both relaxation and pure dephasing  processes occur and the resulting 
decoherence time is  $T_2= [1/(2T_1) + 1/T_2^*]^{-1}$.

Since quantum measurements require averages of measurements runs,
the main effect of fluctuations with $1/f$ spectrum is 
defocusing,  similarly to inhomogeneous  broadening in NMR~\cite{shlichter}.
Fluctuations with large spectral components at low frequencies can be treated 
as stochastic processes in the adiabatic approximation ({\em adiabatic noise}).
The short-times decay of qubit coherences depends on the  symmetry of the 
qubit-environment coupling Hamiltonian. 
For transverse noise, the time dependence is algebraic $\propto [1+ a t^2]^{-1/4}$, 
for longitudinal noise it is exponential quadratic $\propto \exp(- b t^2)$
("static-path"~\cite{PRL05} or "static-noise"~\cite{ithier} approximation).%~\cite{ithier,PRL05}.
 
The simultaneous presence of adiabatic and quantum noise
can be treated in a multi-stage approach~\cite{PRL05}.
In simplest cases, the effects of the two noise components add up 
independently in the coherences time-dependence. Defocusing is minimized 
when noise is transverse with respect to the qubit Hamiltonian~\cite{nak-echo}.
The qubit is said to operate at an "optimal point" 
characterised by algebraic short-times behavior followed by 
exponential decay on a scale $2T_1$.

In the present article we perform a  systematic analysis of the effects and 
interplay of adiabatic and quantum noise on a universal two-qubit gate, extending 
the multi-stage elimination approach introduced in ref.~\cite{PRL05}. 
Understanding these effects is crucial in the perspective of implementing solid-state 
complex  architectures. 
Our system consists of two coupled qubits each affected by transverse and
longitudinal noise with broad-band and non-monotonic spectrum.
Such a general situation has not being studied in the literature.
Previous studies concentrated on harmonic baths with monotonic spectrum  
relying on master equation and/or perturbative Redfield approach~\cite{BlochRedfield},
or on numerical methods~\cite{Thorwart}, or on
formal solutions for selected system observables~\cite{Naegele}.   
 
We quantify entanglement via the concurrence~\cite{concurrence}. To compare 
with bit-wise measurements, single qubit switching probabilities are also evaluated.
Our analysis is based on approximate analytic results and  
exact numerical simulations. 
Our main results are:
(i) The identification of characteristic time scales of entanglement decay due 
to adiabatic noise, quantum noise and their interplay; 
(ii) The characterization of relaxation and dephasing for an entanglement operation 
via the time scales $T_R$, $T_1^{SWAP}$, $T_2^{SWAP}$ and $T_2^{SWAP*}$. 
We point out the dependence of these scales on the symmetry of the Hamiltonian describing 
the interaction between each qubit and the various noise sources; 
(iii) The demonstration that a universal two-qubit gate can be protected against noise 
by operating at an "optimal coupling",  extending the concept of single-qubit "optimal point".

The article is organized as follows. In  Section 2 we introduce the
Hamiltonian model for two-qubit entanglement generation in the presence of
independent noise sources affecting each unit. 
In Section 3 the general features of the power spectra of these fluctuations, 
as observed in single qubit experiments, are summarized. 
The relevant dynamical quantities 
are introduced and the multi-stage approach to eliminate noise variables and 
obtain a reduced description of the two-qubit system is illustrated.  
In Sections 4 and 5 we derive
separately the effect of quantum noise 
within a master equation approach and the leading order effect (quasi-static approximation) 
of adiabatic noise.  
Finally, in Section 6 we  discuss their interplay 
and introduce the relevant time scales characterizing loss of coherence and entanglement
of a universal two-qubit gate in a solid-state environment. Results are  summarized
in \tref{table:decoherence} and in \tref{short}. 
In \ref{appendix-CPB} the entanglement generating model is derived for capacitive
coupled Cooper Pair Boxes (CPBs) including fluctuations of all control parameters.  
In \ref{appendix:SC} the  effect of
selected impurities strongly coupled to the device
is pointed out and we speculate on the possibility to extend the "optimal coupling" scheme
under these conditions.

\section{Universal entangling gate}

Entanglement-generating two qubit gates have been implemented based on different
coupling strategies. In the standard idea of gate-based quantum computation,
the coupling between the qubits is switched on for a quantum gate operation and
switched off after it. The easiest way to realize this scheme is to tune the
qubits in resonance with each other for efficient coupling and move them out of
resonance for decoupling. Employing a fixed coupling scheme 
two-qubit logic gates have been implemented~\cite{coupled-exp-fix}
and high-fidelity Bell states have been generated in 
capacitive coupled phase qubits~\cite{Bell-generation}. A different idea is to
introduce an extra element between the qubits: an adjustable coupler, which can turn the 
coupling on and off~\cite{coupled-th-tunable,coupled-th-adiabatic}.
Alternatively, two-qubit gates are generated by applying microwave signals
of appropriate frequency, amplitude and phase~\cite{coupled-th-ac}. 

The core of the entangling operation of most of the above coupling schemes
consists of two resonant qubits 
with a coupling term transverse with respect to the qubits quantization axis,
as modeled by
\begin{equation}
\mathcal{H}_0 = 
 -\frac{\Omega}{2} \, \sigma_{1z} \otimes \mathbb{I}_{2}
 -\frac{\Omega}{2}  \, \mathbb{I}_{1} \otimes \sigma_{2z}
+ \frac{\omega_c}{2} \, \sigma_{1x} \otimes \sigma_{2x}  \, .
\label{H0}
\end{equation}
Here $\sigma_{\alpha z}$ are Pauli matrices  and
$\mathbb{I}_{\alpha}$ is the identity, in  qubit-$\alpha$ Hilbert space ($\alpha=1,2$).
In our notation, $\sigma_{\alpha z}$ is the qubit-$\alpha$ quantization axis
and we put $\hbar=1$.
This model applies in particular to the fixed, capacitive or inductive, coupling of
superconducting qubits~\cite{coupled-th-fixed}, where
individual-qubit control allows an effective switch on/off of the interaction.

\begin{figure}
\begin{center}
\includegraphics[width=0.8\textwidth]{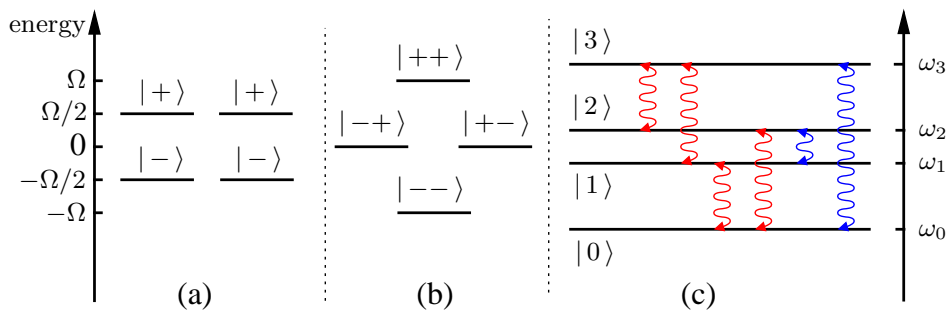}
\end{center}
\caption{(a) Eigenenergies of the uncoupled resonant qubits;
(b) levels in the two-qubit Hilbert space and  logic basis of product states. 
(c) When the coupling is turned on the states $|+-\rangle$ and $|-+\rangle$ mix and 
an energy splitting $\omega_c \ll \Omega$ develops between the 
eigenstates $\{\vert 2 \rangle,\vert 1 \rangle\}$, spanning the 
SWAP subspace. Product states $|--\rangle$ and $|++\rangle$
weakly mix and split, with $\omega_3 - \omega_0= 2 \sqrt{\Omega + (\omega_c/2)^2}$. The
eigenstates $\{\vert 0 \rangle,\vert 3 \rangle\}$ span the Z subspace.
Longitudinal noise in the computational basis is responsible for
inter-doublet relaxation processes (effective transverse inter-doublet)
indicated by red wavy lines (Eq. \eref{H-inter}).
Transverse noise in the computational basis originates incoherent energy exchanges
inside each subspace (effective transverse intra-doublet) indicated by blue wavy lines
(Eqs. \eref{H-trunc-SWAP}, \eref{H-trunc-Z}).
}
\label{splittings}
\end{figure}

Eigenvalues and eigenvectors of eq.~\eref{H0} are reported in  \tref{tab:eigensystem0}.
In order to maintain the single qubit identities, the coupling strength, $\omega_c$, must be
one-to-two orders of magnitudes smaller than the single qubit level spacing, $\Omega$. Thus
eigenvalues form a doublets structure, as schematically illustrated in ~\fref{splittings}.
The Hilbert space factorizes in two subspaces spanned by 
 $\{ |1 \rangle, |2 \rangle\}$ and $\{ |0 \rangle, |3 \rangle\}$.
\begin{table}
\caption{
\label{tab:eigensystem0} 
Eigenvalues and eigenvectors of ${\mathcal{H}}_0 $ expressed in the computational
basis  $\vert \mu \nu \rangle \equiv \vert \mu \rangle_1 \otimes
\vert \nu \rangle_2$,  $\mu,\nu \in \{ +,- \}$ with
$\sigma_{\alpha z} \vert \pm \rangle_\alpha =
\mp \vert \pm \rangle_\alpha$ and $\tan \varphi = - \omega_c/(2 \Omega)$.
}
\begin{indented}
\item[]
\begin{tabular}{ccc}
\br
$i$ & $\omega_i$ & $\vert i \rangle$ \\
\mr
0 & $- \sqrt{\Omega^2 + (\omega_c/2)^2}$ & 
$- ( \sin \varphi/2  )\vert ++ \rangle+ ( \cos \varphi/2 ) \vert -- \rangle$ \\
1 & $-\omega_c/2$ & $(- \vert +- \rangle + \vert -+ \rangle )/\sqrt{2}$ \\
2 & $\omega_c/2$ & $( \vert +- \rangle + \vert -+ \rangle )/\sqrt{2}$ \\
3 & $\sqrt{\Omega^2 + (\omega_c/2)^2}$ & 
$( \cos \varphi/2 )\vert ++ \rangle + ( \sin \varphi/2) 
\vert -- \rangle $ \\
\br
\end{tabular}
\end{indented}
\end{table} 
A couple of qubits described by (\ref{H0}) is suitable to demonstrate entanglement generation.
The system prepared in the factorized state $\vert +- \rangle$ freely evolves
to the entangled state $\vert \psi_e \rangle = [\vert +- \rangle - \rmi \vert -+ \rangle]/\sqrt 2$ in a
time $t_e= \pi /2 \omega_c$, realizing a $\sqrt{{\rm i-SWAP}}$ operation. 
The dynamics takes place inside the $\{ |1 \rangle, |2 \rangle\}$
subspace, which we name "SWAP-subspace". The orthogonal subspace will be instead named
"Z-subspace".

Fluctuations of the control parameters used for the manipulation of individual
qubits couple the circuit to environmental degrees of freedom. 
We consider the general situation where each qubit is affected 
both by longitudinal noise (coupled to $\sigma_{\alpha z}$) and 
by transverse noise (coupled to $\sigma_{\alpha x}$), as described by the interaction Hamiltonian
\begin{equation}
\label{eq:Hnoise}
\mathcal{H}_\mathrm{I} = -\frac{1}{2} \left [ \hat{x}_{1} \, \sigma_{1x}
 +  \hat{z}_{1} \, \sigma_{1z} \right ]
\otimes \mathbb{I}_{2}
-
\frac{1}{2} \mathbb{I}_{1} \otimes \left [ 
\hat{x}_{2} \, \sigma_{2x}+ \hat{z}_{2} \, \sigma_{2z} \right ] \,.
\end{equation}
Here $\hat{x}_{\alpha}$ and $\hat{z}_\alpha $ are collective environmental quantum
variables coupled to different qubits degrees of freedom.
For instance, in the case of two CPB-based qubits at the charge optimal point~\cite{vion}, 
the charge operator is  $\sigma_{\alpha x}$, and the Josephson 
operator is   $\sigma_{\alpha z}$.
Fluctuations of the gate charge are described by a transverse coupling term, 
$\hat{x}_{\alpha} \, \sigma_{\alpha x}$, and 
noise in the superconducting phase by the longitudinal term, 
$\hat{z}_{\alpha} \, \sigma_{\alpha z}$ ~\cite{ithier} (see \ref{appendix-CPB} for the derivation)
~\footnote{In our notation the qubit's quantization axis is $\sigma_{\alpha z}$,
irrespective of the working point. Thus,  $\hat{x}_{\alpha}$ and $\hat{z}_\alpha $ 
describe physically different processes only at selected operating points. Usually, this
is the case at the single qubit's optimal points, see  \ref{appendix-CPB}.}. 
The complete device Hamiltonian reads  $\mathcal{H}_0 + \mathcal{H}_\mathrm{I} + \mathcal{H}_\mathrm{R}$,
where $\mathcal{H}_\mathrm{R}$ denotes the free Hamiltonian of all environmental variables.

In order to identify relaxation and pure dephasing processes for the coupled
qubit setup we project $\mathcal{H}_0+ \mathcal{H}_\mathrm{I}$ 
in the 4-dim Hilbert space generated by the eigenstates of $\mathcal{H}_0$,  
$\{ |i \rangle \}$, $i=0,1,2,3$, where we may rewrite 
\begin{eqnarray}
\label{H0-proj}
\!\!\!\!\!\!\!\!\!\!\!\!\!\!\!
\mathcal{H}_0 &= & \sum_i \omega_i |i \rangle \langle i| =
\frac{\omega_c}{2} \, \big [|2\rangle \langle 2| - | 1\rangle \langle 1| \big ]  \, +
 \, \sqrt{\Omega^2 + (\omega_c/2)^2} \, \big [|3\rangle \langle 3| - | 0\rangle \langle 0| \big ] \\
\!\!\!\!\!\!\!\!\!\!\!\!\!\!\!
 \mathcal{H}_\mathrm{I} &=& \frac{1}{2} (\hat{x}_{1} + \hat{x}_{2}) \, \Big [A_-  |2\rangle \langle 0| 
 + A_+ |2\rangle \langle 3| + {\rm
 h.c.} \Big ] \, +
 \nonumber \\ &+& 
 \frac{1}{2} (\hat{x}_{1} - \hat{x}_{2})  \, \Big [- A_+  |1\rangle \langle 0| + A_- | 1\rangle \langle 3| + {
 \rm h.c.} \Big ]  \nonumber\\
 &-&  \frac{1}{2} (\hat{z}_{1} - \hat{z}_{2})  \,  \Big [|1\rangle \langle 2| + |2\rangle \langle 1|\Big ] 
 \label{HI-proj}\\
 &-& \frac{1}{2} (\hat{z}_{1} + \hat{z}_{2})  \, \Big [ \, \cos \varphi  \, ( |0\rangle \langle 0| - | 3\rangle \langle 3|)  
 + \sin \varphi \,  ( |0\rangle \langle 3| + | 3\rangle \langle 0|)\, \Big ]  \nonumber
\end{eqnarray}
where  $A_\pm = [\cos(\varphi/2) \pm \sin(\varphi/2)]/\sqrt 2$, with 
$\tan \varphi = - \omega_c/(2 \Omega)$.
In (\ref{HI-proj}) we distinguish "effective longitudinal"  
and "effective transverse" terms. The first ones are diagonal
in the eigenbasis $\{ |i \rangle \}$
and are responsible for pure dephasing processes.
"Effective transverse" terms instead are off-diagonal and originate 
both intra- and inter-doublet relaxation processes.
Specifically we have:\\ 
 {\bf SWAP subspace $\{ |1\rangle , |2 \rangle\}$}:  Longitudinal noise in the computational basis 
$\propto \sigma_{\alpha z} \hat{z}_{\alpha}$,
 originates "effective transverse" noise in the SWAP-subspace, i.e. the restriction of 
 $\mathcal{H}_0+ \mathcal{H}_\mathrm{I}$ to $\{ |1\rangle , |2 \rangle\}$ reads
 \begin{equation}
 \mathcal{H}_{SWAP}^{proj} = \frac{\omega_c}{2} \, \big [|2\rangle \langle 2| - | 1\rangle \langle 1| \big ] -
 \frac{1}{2}(\hat{z}_{1} - \hat{z}_{2} ) \,  \Big [|1\rangle \langle 2| + |2\rangle \langle 1|\Big ] 
 \label{H-trunc-SWAP}
 \end{equation}
 Note that if both qubits were affected by the same longitudinal noise 
  no effective transverse noise in 
 this subspace would be present. This situation may occur in the presence of 
 totally correlated noise affecting  both qubits~\cite{NJP-special}.\\
 {\bf Z subspace $\{ |0 \rangle , |3 \rangle\}$}: Longitudinal noise in the computational basis 
 originates both  "effective transverse" and  "effective longitudinal" noise in the Z-subspace, 
 i.e. the projection of 
 $\mathcal{H}_0+ \mathcal{H}_\mathrm{I}$ on $\{ |0\rangle , |3 \rangle\}$ reads
 \begin{eqnarray}
 \mathcal{H}_Z^{proj} &=& \sqrt{\Omega^2 + (\omega_c/2)^2} \, 
\big [|3\rangle \langle 3| - | 0\rangle \langle 0| \big ] 
\nonumber \\
 &+& (\hat{z}_{1} + \hat{z}_{2} ) \, 
\Big [ \, \cos \varphi  \, ( |0\rangle \langle 0| - | 3\rangle \langle 3|)  
 + \sin \varphi \,  ( |0\rangle \langle 3| + | 3\rangle \langle 0|)\, \Big ] \, .
 \label{H-trunc-Z}
 \end{eqnarray}
 Effective longitudinal and transverse components are modulated via the mixing angle, 
 $\varphi$,  similarly to a single qubit with operating point $\varphi$.\\
{\bf Inter-doublet processes}: The only effect of transverse noise in the computational 
basis is to mix the two subspaces via "effective transverse" inter-doublet
terms (Fig.\ref{splittings} (c)) 
\begin{eqnarray}
\mathcal{H}_{inter} &=& \frac{1}{2}(\hat{x}_{1} + \hat{x}_{2}) \, \Big [A_-  |2\rangle \langle 0| 
 + A_+ |2\rangle \langle 3| +{\rm h.c.} \Big ] \nonumber \\  
  \, &+& \, \frac{1}{2}(\hat{x}_{1} - \hat{x}_{2})  \, \Big [- A_+  |1\rangle \langle 0| + A_- | 1\rangle \langle 3| + {
 \rm h.c.} \Big ]
 \label{H-inter}
\end{eqnarray}
These inter-doublet terms are responsible, in particular, for relaxation processes from 
the SWAP-subspace to the ground state. We will demonstrate that the resulting "global" relaxation time
sets the upper limit to all other gate operation times, including other
decoherence time scales.

\section{Multi-scale approach to broad-band noise}

The considered entanglement generating operation takes place in the presence
of broad-band and non-monotonic noise. In this Section we review the
multi-stage elimination approach to deal with this problem. The method has been introduced for a
single qubit in ref.~\cite{PRL05} where the various approximations have been
checked by comparing with the exact numerical solution of the system evolution.
This approach allowed to accurately explain the observed dynamics in different
experiments~\cite{ithier,Bylander11}, confirming its appropriateness to deal with the more
complex system studied in the present article. 
The method has been extended to a multi-qubit gate in ref.~\cite{PhysScript09}, 
here we summarize the main steps.

\begin{figure}[t!]
\centering
\includegraphics[width=0.5\textwidth]{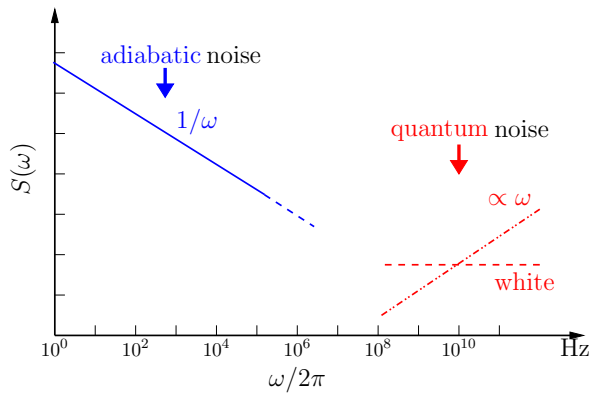}
\caption{Sketch of the typical power spectrum of the environmental variable $\hat E_\alpha$
(logarithmic scale). 
Measurements of $1/f$ noise usually extend between $1$~Hz and $0.1 - 1$~MHz, whereas
the ohmic or white spectrum region typically ranges around $5-20$~GHz~\cite{ithier,Bylander11}.
The region classified as adiabatic noise and quantum noise are indicated.
}
\label{fig:spectra}
\end{figure}

The multi-stage elimination approach is based on a classification of the noise sources according 
to their effects and circumvents the problem of a microscopic description of noise sources, 
which are often non Gaussian and non Markovian~\cite{PRL02,Altshuler}.
In this perspective, the required statistical information on the environment depends on the 
specific quantum operation performed and on the measurement protocol.
Even if the statistical characterization of the environment 
requires going beyond the second order cumulant, often knowledge 
of the power spectrum of the bath variables, here
$ \hat x_\alpha$ and $ \hat z_\alpha$ denoted generically as $\hat{E}_{\alpha}$, 
\begin{equation}
 S_{E_\alpha}(\omega)\,=\,\frac{1}{2} \, \int_{0}^{+\infty} \rmd t \, \rme^{-\rmi \,  \omega t}\, 
\, [\,\langle\langle \hat E_\alpha(t) \hat E_\alpha(0) \rangle \rangle
+ \langle\langle \hat E_\alpha(0) \hat E_\alpha(t) \rangle \rangle
\,] \,,
\label{SEalpha} 
\end{equation}
is sufficient. Where
$\langle \langle \dots \rangle \rangle$ denotes the equilibrium average with respect 
to $\mathcal{H}_\mathrm{R}$  and we assumed stationary processes 
with $\langle \langle \hat E_\alpha \rangle \rangle =0$.
The typical power spectrum reported in various single qubit experiments is sketched 
in \fref{fig:spectra}. To make explicit reference to some practical situations, 
in \tref{tab:noise1} we summarize the characteristics of transverse and longitudinal noise spectra
at low- and high-frequencies reported in a CPB-based circuit~\cite{ithier} and in
the recent experiment on a flux qubit~\cite{Bylander11}.  

The low-frequency part of the spectra of each variable is $1/f$, 
$S_{E_\alpha}^{1/f}(\omega)= A_{E_\alpha}/\omega$. 
The amplitude $A_{E_\alpha}$ can be estimated from  spectral measurements.
If  $\gamma_m$ and  $\gamma_M$ denote respectively the low- and high-frequency cut-offs
of the $1/f$ region, then 
$A_{E_\alpha}=\pi \sigma_{E_\alpha}^2  [\ln(\gamma_M/\gamma_m)]^{-1}$, where
$\sigma_{E_\alpha}^2$ is the variance, 
$\sigma_{E_\alpha}^2 = \int_{\gamma_m}^{\gamma_M} \frac{d \omega}{\pi} S^{1/f}_{E_\alpha}(\omega)$.
It  can be approximated as
$\sigma_{E_\alpha}^2 = \int_{1/t_m}^{\gamma_M} \frac{d \omega}{\pi} S^{1/f}_{E_\alpha}(\omega)$ 
where $t_m$ is the overall acquisition time for a single data point which results from averaging
over several measurement trials.
\footnote{The intrinsic high-frequency cut-off of the $1/f$ spectrum depends on the specific 
microscopic source and it is usually not detectable in experiments. In the present article
we discuss measurements protocols where the details of the behavior
of the power spectrum below $\gamma_m$ and around $\gamma_M$ are not relevant and
results depend logarithmically on the ratio $\gamma_M/\gamma_m$.  }

The high-frequency power spectrum is usually inferred indirectly from measurements of the
qubit relaxation times under various protocols~\cite{nak-spectrum,ithier,Bylander11}.
From the resulting figures the expected spectra for the
corresponding quantum variables $\hat E_\alpha$ 
at the relevant frequencies, $\Omega / 2 \pi$ and $\omega_c / 2 \pi$ are derived.
Experiments tuning the single qubit level spacing $\Omega$ reveal either ohmic
~\cite{nak-spectrum,Bylander11} or white~\cite{ithier} power spectrum in the GHz range.
Evidence of spurious resonances in the spectrum have often been reported, they show up
as beatings in time resolved  measurements~\cite{vion,Ustinov}.
\begin{table}[t!]
\caption{
\label{tab:noise1}
Characteristics of transverse and longitudinal noise at low- and high-frequencies inferred 
from data reported in ref.\cite{vion} for a charge-phase qubit at its double optimal 
point with $\Omega \approx 2 \pi  \times 16$GHz 
(charge noise is transverse and phase noise is longitudinal),
and in ref.\cite{Bylander11} for a flux qubit at the optimal point with 
$\Omega \approx 2 \pi \times 5$GHz
(flux noise is transverse and critical current noise is longitudinal). For the charge-phase
qubit we considered $\gamma_M/\gamma_m = 10^6$, results logarithmically  depend on this ratio.
}
\begin{indented}
\item[]\begin{tabular}{ccc}
\br
& Charge-Phase Qubit & Flux Qubit  \\
\mr
%\hline
$S_x^{1/f}$ &  $ \sigma_x \approx 2 \times 10^{-2} \,\Omega $ & $ \sigma_x \approx 2 \times 10^{-3} \,\Omega$ \\
$S_z^{1/f}$  &  $ \sigma_z \approx 10^{-6} \,\Omega $ & $ \sigma_z \approx 10^{-5} \,\Omega$ \\
\hline
$S_x$   & $S_{x}(\Omega) \approx 4 \times 10^6\,$s$^{-1}$ & $S_{x}(\Omega) \approx 2 \times 10^5\,$s$^{-1}$ \\
$S_z$  & $S_{z}(\Omega) \approx 10^5$ s$^{-1}$    &                          -                      \\
%\hline
\br
\end{tabular}
\end{indented}
\end{table}

In our analysis noise sources belong to three classes.
Low frequency noise with $1/f$ spectrum is adiabatic since it does not induce 
transitions, but mainly defocuses the signal, we classify it as {\em adiabatic noise}.
Noise at frequencies of the order of the qubits splittings 
is responsible for dissipation and ultimately for spontaneous decay, thus it is classified as
{\em quantum noise}.
Possible resonances in the spectrum pertain to the class named {\em strongly coupled noise}.
Noise sources belonging to different classes act on different frequency scales and are
treated via specific approximation schemes.
The distiction can be illustrated as follows.
We are interested to a reduced description of the $2$-qubit system, 
expressed by the reduced density matrix (RDM),  $\rho(t)$ 
obtained by tracing out environmental degrees of freedom from the total density matrix 
$\rho^{\rm Q,A,SC}(t)$, which depends on  quantum (Q), adiabatic (A) and strongly coupled 
(SC) bath variables. Since bath's degrees of freedom belonging to different classes of noise
act of different time scales we separate in each part of the interaction Hamiltonian, 
$\sigma_{\alpha i} \,\hat E_{\alpha}$
($i=x,z$), the contribution from various noise classes as follows 
\begin{equation}
\sigma_{\alpha i } \, \hat E_{\alpha} \; \to \; \sigma_{\alpha i} \, \hat E_{\alpha}^{\rm Q} + 
\sigma_{\alpha i} \, \hat E_{\alpha}^{\rm A} + \sigma_{\alpha i} \, \hat E^{\rm SC}_{\alpha} \, .
\label{int-plit}
\end{equation}
Adiabatic noise, $\hat E_{\alpha}^{\rm A}$, is typically correlated on a time scale much longer than 
the inverse of the qubit's frequencies, $\Omega_\alpha$, then in the spirit 
of the Born-Oppenheimer approximation it can be seen as a classical stochastic field  
$ \{E_{\alpha}(t)\} \equiv  \vec E(t)$.
This approach is valid when the contribution of adiabatic noise to 
spontaneous decay is negligible, a necessary condition being 
$t \ll T_1^A \propto S_{E_A}(\Omega_\alpha)^{-1}$. This condition is usually satisfied at
short enough times, 
since $S_{E_A}^{1/f}(\omega)$ is substantially  different from zero only at frequencies 
$\omega \ll \Omega_\alpha$.
 This fact suggests how to trace-out different noise classes
in the appropriate order.  The total density matrix parametrically depends on 
the specific realization of the slow random drives $\vec E(t)$ 
and may be written as $\rho^{\rm Q,A,SC}(t)= \rho^{\rm Q,SC}(t | \vec E(t))$. 
The first step is to trace out quantum noise. In the simplest cases this requires
solving a master equation. In a second stage, the average over all the realizations of
the stochastic processes, $\vec E(t)$, is performed. This leads to a reduced density
matrix for the $2$-qubit system plus the strongly coupled degrees of freedom.
These latter have to be traced out in a final stage  by solving the Heisenberg
equations of motion, or by approaches suitable to the specific microscopic 
Hamiltonian or interaction. For instance, the dynamics may be solved exactly
for some special quantum impurity models at pure dephasing, 
when impurities are longitudinally coupled to each qubit~\cite{PRL02,Altshuler}.  
The multi-stage elimination  can be formally written as
\begin{equation}
\rho(t) = Tr_{\rm SC} \left \{ 
\int {\mathcal D}[\vec E(t)] \,  P[\vec E(t)] \; 
Tr_{\rm Q} \Big[ \, \rho^{\rm Q,SC}\Big(t | \vec E(t)\Big) \, \Big]\right\} \, ,
\end{equation}
where $Tr_{\rm Q}$ and $Tr_{\rm SC}$ indicate respectively the trace over the Q and
SC degrees of freedom.
In the following Sections we apply the multi-stage approach to the two-qubit
gate by tracing out first quantum noise and secondly the adiabatic noise.
In \ref{appendix:SC} the effect of a SC impurity will be analyzed.

\subsection{Relevant dynamical quantities}

We focus on the $\sqrt{{\rm i-SWAP}}$ operation  
$| +- \rangle \to |\psi_e \rangle = [| +- \rangle - i | -+ \rangle]/\sqrt{2}$
which generates  by free evolution an entangled state
at $t_e = \pi/2 \omega_c$.   
As a unambiguous test of entanglement generation and its degradation due to noise, 
we calculate the evolution of concurrence during the gate operation. 
Introduced in ref.~\cite{concurrence}, the concurrence quantifies the 
entanglement of a pair of qubits, being $C=0$ for separable
states and $C=1$ for maximally entangled states. 
For the situations discussed in the present article and specified in the following
Sections, the two-qubits RDM takes the "X-form", i.e. 
the RDM expressed in the eigenstates basis is non-vanishing only along the diagonal 
and anti-diagonal at any time. Under these conditions, the concurrence can be evaluated 
in analytic form~\cite{Xform}. In general, it depends both on diagonal and off-diagonal elements
of the RDM.

In order to directly compare with experiments where bit-wise readout is performed,
we also evaluate 
the qubit 1 switching probability $P_{\mathrm{SW}1}(t)$, 
i.e. the probability that it will pass to the state $\vert - \rangle_1$ 
starting from the state $\vert + \rangle_1$; 
and the probability $P_2(t)$ of finding the qubit 2 
in the initial state $\vert - \rangle_2$. In terms of the two qubit RDM 
in the eigenstate basis they read (${\rm Tr}_i \rho(t)$ denotes
partial trace over qubit $i$ of the two-qubit density matrix)
\begin{eqnarray}
P_{\mathrm{SW}1}(t)  &=& \, _1 \! \langle - | {\rm Tr}_2 \rho(t) | - \rangle_1 
 = \frac{1}{2} \left [ \, \rho_{11}(t) + \rho_{22}(t) \right ]
+ \rho_{00}(t)  \nonumber \\ 
&+& \left [ \, \rho_{33}(t) - \rho_{00}(t) \right ] \sin^2 \frac{\varphi}{2} 
+  {\rm Re}[\rho_{12}(t)] +  {\rm Re} [\rho_{03}(t)] \sin \varphi
\label{eq:psw1} \\
P_2(t) &=& _2 \! \langle - | {\rm Tr}_1 \rho(t) | - \rangle_2 
=
\frac{1}{2} \left [ \, \rho_{11}(t) + \rho_{22}(t) \right ] 
+  \rho_{00}(t)  \nonumber \\ 
&+&  \left [ \, \rho_{33}(t) - \rho_{00}(t) \right ] \sin^2 \frac{\varphi}{2} 
-  {\rm Re}[\rho_{12}(t)] + {\rm Re}[\rho_{03}(t)] \sin \varphi.
\label{eq:psw2}
\end{eqnarray}
For preparation in the state $| +- \rangle$, in the absence of external fluctuations
the above probabilities  read
\begin{equation}
P_{\mathrm{SW}1}(t) =  \frac{1- \cos \omega_c t}{2} \; \quad , \quad \;
P_2(t) = \frac{1+ \cos \omega_c t}{2} \,.
\label{eqs:switching}
\end{equation}
The cyclic anti-correlation of the probabilities signals the formation 
of the entangled state, as reported in various recent
experiments~\cite{coupled-exp-fix,coupled-exp,Bell-generation,nguyen}.

Both the concurrence and the switching probabilities depend on combinations of
populations and coherences in the eigenbasis. 
Therefore the relevant time scale to quantify the "quality factor" or the efficiency of
the universal two-qubit gate is not simply related to a specific RDM element, as a difference 
with a single qubit gate, where the decay time of the qubit 
coherence quantifies the quality factor of the operation (with $T_2$ due partly to relaxation 
processes ($2 T_1$), partly to Markovian pure dephasing processes $T_2^*$ or  
originated from inhomogeneous broadening).
In the following we will study the time dependence of coherences and populations, and
identify the environmental processes (transverse, longitudinal, low frequency, 
high frequency, etc.)
which originate various decay times. Based on this analysis we will discuss 
the resulting effect on the decay of the switching probabilities and of the concurrence.

\section{Quantum noise}
 \label{sec:quantumnoise}
To begin with, we consider  the effect of quantum noise replacing in \eref{int-plit} 
 $\hat{E}_{\alpha} \to  \hat E_{\alpha}^{\rm Q}$.
The system dynamics is obtained by solving the Born-Markov master equation
for the RDM. In the system eigenstate basis and performing
the secular approximation (to be self-consistently checked) it takes the
standard form~\cite{cohen,PhysicaE}:
\begin{eqnarray}
\dot \rho_{ii}(t) &=& - \sum_{m \neq i} \Gamma_{im} \, \rho_{ii}(t) + 
\sum_{m \neq i} \Gamma_{mi} \, \rho_{mm}(t) 
\label{eq:populations}
\\
\dot \rho_{ij}(t) &=& - (i \tilde \omega_{ij} + \widetilde\Gamma_{ij} ) 
\,\rho_{ij}(t) \,.
\label{eq:coherences}
\end{eqnarray}
The rates $\Gamma_{im}$, $\tilde\Gamma_{ij}$ and the frequency shifts 
$\tilde \omega_{ij} -  \omega_{ij}$,
where $ \omega_{ij}= \omega_i -\omega_j$, depend respectively on the real
and imaginary parts of the lesser and greater Green's functions
which describe  emission (absorption) rates to (from)  the reservoirs
\begin{eqnarray}
\int_0^\infty d t e^{i \omega t} 
\langle \langle \hat E_{\alpha}(t) \hat E_{\alpha}(0) \rangle \rangle 
&=& \frac{1}{2}  C_{E_\alpha}(\omega) -  \frac{i}{2} {\mathcal E}_{E_\alpha} (\omega)\\
\int_0^\infty d t e^{i \omega t} 
\langle \langle \hat E_{\alpha}(0) \hat E_{\alpha}(t)  \rangle \rangle &=&
 \frac{1}{2} C_{E_\alpha}(- \omega) + \frac{i}{2} {\mathcal E}_{E_\alpha} (- \omega) \, .
\end{eqnarray}
In terms of the corresponding power spectra they read 
\begin{eqnarray}
C_{E_\alpha}(\omega) &=& \frac{2 \, S_{E_\alpha}(\omega)}{1+ \exp{(- \omega/ k_B T)}} 
\label{absorption}\\
 {\mathcal E}_{E_\alpha} ( \omega) &=& {\mathcal P} \int_{-\infty}^{\infty} \frac{d \omega^\prime}{2 \pi} \, 
\frac{C_{E_\alpha}(\omega^\prime)}{\omega^\prime -\omega}
\label{principalv} \,,
\end{eqnarray}
where ${\mathcal P}$ denotes the principal value of the integral.
Due to the symmetry of $\mathcal{H}_0 + \mathcal{H}_\mathrm{I}$, 
Eqs. \eref{H0-proj} - \eref{HI-proj}, the only 
independent emission rates, $\Gamma_{ij}$, are $\Gamma_{10}=\Gamma_{32}$,
$\Gamma_{20}=\Gamma_{31}$, $\Gamma_{21}$, $\Gamma_{30}$, see \fref{splittings}.
Symmetric relations hold between the corresponding absorption
rates $\Gamma_{ji}$.  These processes originate from "effective transverse" noise, in particular
transverse fluctuations ($\propto \hat x_{\alpha}$) enter the rates
connecting the SWAP and the Z subspaces, whereas longitudinal
fluctuations  ($\propto \hat z_{\alpha}$) enter the intra-subspace rates, 
$\Gamma_{21}$, $\Gamma_{30}$, cfr eqs.~\eref{HI-proj}~-~\eref{H-inter}.
They read
\begin{eqnarray}
\begin{array}{ll}
\label{gamma10}
\Gamma_{10} = \frac{1}{8} \, (1+\sin \varphi) \, [ C_{x_1}(\omega_{10}) 
+  C_{x_2}(\omega_{10})] \qquad \, {\rm  inter-subspace}
\\
\label{gamma20}
\Gamma_{20} = \frac{1}{8} \, (1-\sin \varphi) \, [ C_{x_1}(\omega_{20}) 
+  C_{x_2}(\omega_{20})]  \qquad \, {\rm  inter-subspace}
\\
\label{gamma30}
\Gamma_{30} = \frac{1}{4} \, \sin^2 \varphi \, [C_{z_1}(\omega_{30}) 
+ C_{z_2}(\omega_{30})]  \qquad  \qquad \, {\rm  intra-subspace}
\\
\label{gamma21}
\Gamma_{21} = \frac{1}{4} \,  [C_{z_1}(\omega_{21}) 
+ C_{z_2}(\omega_{21})]   \qquad \qquad  \qquad  \; \; \;{\rm intra-subspace}
\end{array}
\label{eq:ratesgeneral}
\end{eqnarray}
Absorption rates have the same form with $C_{E_\alpha}(\omega_{lm})$
replaced by  $ C_{E_\alpha}(-\omega_{lm})$.
The imaginary parts of the corresponding terms take similar forms. 
They enter the frequency shifts as reported in \ref{appendix:shifts}.

In the secular approximation, the SWAP and Z coherences decay exponentially  
with rates
\begin{eqnarray}
\widetilde \Gamma_{12} &=& \frac{1}{2} \, [\Gamma_{10} + \Gamma_{01} + \Gamma_{20} + \Gamma_{02} 
+ \Gamma_{12} + \Gamma_{21}] 
\label{gamma12tilde}\\
\widetilde \Gamma_{30} &=& \frac{1}{2} \, [ \Gamma_{10} + \Gamma_{01}
 + \Gamma_{20} + \Gamma_{02} + \Gamma_{30} + \Gamma_{03}] + \Gamma^*_Z\, ,
\end{eqnarray}
both inter-subspace and intra-subspace rates enter the decay of the SWAP and Z coherences.
Note that the decay rate of the coherence $\rho_{12}(t)$ is only originated from dissipative
processes (intra- or inter- effective transverse)
since no pure dephasing processes (effective longitudinal noise) inside the SWAP 
subspace exist, cfr eq.~\eref{HI-proj}. 
On the contrary, the coherences in the Z subspace also decay because of the effective longitudinal 
terms $\cos \varphi ( |0\rangle \langle 0| - | 3\rangle \langle 3|)
(\hat{z}_{1} + \hat{z}_{2})$, which originate 
$ \Gamma^{Z*} = \frac{1}{4} \cos^2 \varphi [S_{z1}(0)+S_{z2}(0)]$.
This pure dephasing factor adds up to a decoherence rate due to
intra-subspace effective transverse noise having the characteristic
form $[T_2^Z]^{-1} = (\Gamma_{30}+ \Gamma_{03})/2$,
as implied by \eref{H-trunc-Z}, and to inter-doublet relaxation rates.

Equations \eref{eq:populations} for the populations do not
decouple even in the secular limit. General solutions are quite
cumbersome, so here we report expressions in the small temperature
limit with respect to the uncoupled qubits splittings, 
$k_B T \ll \Omega$.
In this regime,  if the system is initially
prepared in the state  $| +- \rangle = (|2 \rangle - |1 \rangle)/\sqrt{2}$, 
level $3$ is not populated, $\rho_{33}(t)=0$, and  the Z-coherences vanish, 
$\rho_{03}(t) = \rho_{03}(0) = 0$. The remaining populations
are conveniently expressed in terms of the escape rates from 
levels $1$ and $2$
\begin{equation}
\Gamma_1^{\rm e} = \Gamma_{10} + \Gamma_{12} \quad \, , \, \quad
\Gamma_2^{\rm e} = \Gamma_{20} + \Gamma_{21} \,,
\label{eq:escape}
\end{equation}  
which enter the evolution of the populations in the following combinations
\begin{eqnarray}
\Gamma_{\pm} &=&  \frac{\Gamma_1^{\rm e} + \Gamma_2^{\rm e}}{2}
\mp \frac{1}{2} \, \sqrt{(\Gamma_1^{\rm e} - \Gamma_2^{\rm e})^2 
+ 4 \Gamma_{12} \Gamma_{21}} \, .
\label{eq:rates}
\end{eqnarray}
For the chosen initial conditions the populations read
\begin{eqnarray}
\rho_{11}(t) &=& \frac{\Gamma_{21}}{2(\Gamma_- - \Gamma_+)} 
%\left \{ 
\sum_{k=\pm} k
\left [1 + \frac{\Gamma_{12}}{\Gamma_1^{\rm e}- \Gamma_k} \right ] \,
 e^{-\Gamma_k t}
\label{population1}
\\
\rho_{22}(t) &=& \frac{1}{2(\Gamma_- - \Gamma_+)} 
%\left \{ 
\sum_{k=\pm} k
\left [\Gamma_{12} - \Gamma_k + \Gamma_1^{\rm e} \right ] \,
 e^{-\Gamma_k t} 
\label{population2}
\\
\label{population0}
 \rho_{00}(t) &=& 1 -(\rho_{11}(t) + \rho_{22}(t)) \, .
\end{eqnarray}
Note that, since $k_B T \lesssim \omega_c \ll \Omega$,  thermal
excitation processes internal to the SWAP subspace, expressed
via the absorption rate $\Gamma_{12}$, cannot be neglected
and $\Gamma_{21} \approx \Gamma_{12}$.
On the contrary, inter-doublet thermal excitation processes are 
exponentially suppressed with respect to the corresponding decay rates,
$\Gamma_{01} \, , \Gamma_{02} \, \ll  \,\Gamma_{10} \, , \Gamma_{20}$, with 
$\Gamma_{10} \approx \Gamma_{20}$.
Thus the SWAP coherences decay rate \eref{gamma12tilde} is approximately
half  the sum of the escape rates (\ref{eq:escape})  
\begin{equation}
\label{tildeGamma12}
\widetilde \Gamma_{12} \approx \frac{\Gamma_1^{\rm e} + \Gamma_2^{\rm e}}{2} \, .
\end{equation}
In order to observe generation of entanglement inside the SWAP subspace it is 
necessary that relaxation processes to the ground state take place on 
a sufficiently long time scale. 
This is guaranteed when inter- and intra-subspace rates  satisfy the condition
$\Gamma_{10}, \Gamma_{20} \ll \Gamma_{21} , \Gamma_{12}$,
which requires that the spectra of the originally 
transverse and longitudinal fluctuations are  
$S_{x_\alpha}(\Omega) \ll S_{z_\alpha}(\omega_c)$ (from \eref{eq:ratesgeneral}). 
In this regime, the SWAP decoherence rate \eref{tildeGamma12} is due 
to effective transverse processes internal to the subspace
\begin{equation}
\label{gamma12-appr}
\widetilde \Gamma_{12} \approx \frac{1}{2} [ \Gamma_{12} + \Gamma_{21}] \, ,
\end{equation} 
and the scales entering the populations take the approximate forms
\begin{eqnarray}
\label{gammaplus}
\Gamma_+  &\approx &  \frac{1}{2} [ \Gamma_{10} + \Gamma_{20}] \qquad 
\qquad {\rm relaxation \, to \, the \, ground \,state} \\
\label{gammaminus}
\Gamma_-  &\approx &  \Gamma_{12} + \Gamma_{21} %= 2 \widetilde \Gamma_{12} 
\qquad \qquad \quad \, 
{\rm relaxation \, inside \, the \, SWAP \, subspace}
\end{eqnarray}
with $\Gamma_+ \ll \Gamma_-$.  
Therefore, the time scales resulting from quantum noise, considering that
$k_B T \lesssim \omega_c \ll \Omega$, when 
$S_{x_\alpha}(\Omega) \ll S_{z_\alpha}(\omega_c)$, are 
\begin{itemize}
\item "Global" relaxation time to the ground state, analogous to the single qubit $T_1$:
Its order of magnitude is 
the spectrum of {\em transverse fluctuations} in the computational basis at frequency $\Omega$
\begin{equation}
T_R = 1/ \Gamma_+ \approx \frac{8}{C_{x_1}(\Omega)+C_{x_2}(\Omega)} =  \frac{4}{S_{x_1}(\Omega)+S_{x_2}(\Omega)} \, ,
\label{T_one}
\end{equation}
where the approximate form comes from  eq.~(\ref{eq:ratesgeneral}), 
where $\omega_{10} \approx \omega_{20} \approx \Omega$. 
\item Relaxation/decoherence times inside the SWAP subspace: 
They are due to "effective transverse" fluctuations
inside this subspace, physically originated from {\em longitudinal noise} on each qubit
at frequency $\omega_c$. 
Since there is no effective longitudinal noise in the SWAP subspace, relaxation and
dephasing times are related by the typical relation $T^{SWAP}_2 \approx 2 T^{SWAP}_1$
where
\begin{eqnarray}
T^{SWAP}_1 &=&1/\Gamma_- \approx 1/ (2 \widetilde \Gamma_{12}) 
\label{T1swap}\\
T^{SWAP}_2 &=& 1/\widetilde \Gamma_{12}  
\approx \frac{4}{S_{z1}(\omega_c) + S_{z2}(\omega_c)} \, .
\label{T2swap}
\end{eqnarray}
\end{itemize}
We now briefly comment on the validity of the secular approximation. 
It consists in separating  the evolutions of elements $\rho_{ij}(t)$ and 
$\rho_{lm}(t)$ provided that $|\omega_{ij} - \omega_{lm}|  \gg \tau^{-1}$, 
where $\tau$ denotes the typical evolution time scale of the system ~\cite{cohen}. 
In the  present case this condition is fulfilled if 
$\omega_{21} \approx \omega_c \gg \widetilde \Gamma_{12}$.
This constraint, on the other side, {\em  needs} to be satisfied in order to
observe generation of entanglement in the presence of quantum noise, 
as expressed for instance from anti-correlation of the probabilities
(\ref{eq:psw1}) and (\ref{eq:psw2}).
Thus it can be regarded as a 
{\em necessary condition}, whose validity  has to be checked
case by case and requires (from \eref{gamma12tilde})
\begin{equation}
S_{x_\alpha}(\Omega) \, , \, S_{z_\alpha}(\omega_c) \ll  \omega_c \, .
\label{wc-cond}
\end{equation}
In conclusion, entanglement generation in the
presence of transverse and longitudinal quantum noise is guaranteed when
\begin{eqnarray}
&& S_{x_\alpha}(\Omega) \ll S_{z_\alpha}(\omega_c) 
\label{strong}\\
&& \widetilde \Gamma_{12} \approx \frac{S_{z1}(\omega_c) + S_{z2}(\omega_c)}{4}  \ll  \omega_c   \, .
\end{eqnarray}
Under these conditions, the "global" relaxation time and the SWAP relaxation and decoherence
times are given respectively by eqs. \eref{T_one} and \eref{T1swap}, \eref{T2swap}.
We note that, as a limiting case,  the $\sqrt{{\rm i-SWAP}}$ operation can be realized
also releasing the condition $\Gamma_+ \ll \Gamma_-$,
 provided both rates are much smaller than the coupling strength $\omega_c$,
and $\widetilde \Gamma_{12} \ll \omega_c$.
For instance, 
when $(\omega_c/ \Omega) S_{x_\alpha}(\Omega) \ll S_{z_\alpha}(\omega_c)\ll
\omega_c$  in Eq. \eref{eq:rates}
$|\Gamma_{10} - \Gamma_{20}| \ll \Gamma_{21} + \Gamma_{12}$ and
the rates $\Gamma_\pm$ are still given by Eqs. \eref{gammaplus}, \eref{gammaminus} 
leading to $T_R = 1/ \Gamma_+$ and $T^{SWAP}_1 =1/\Gamma_-$.
The SWAP dephasing time in this case also depends on trasverse fluctuations, 
$T^{SWAP}_2 = 1/\widetilde \Gamma_{21} \approx 1/[\Gamma_+ + \Gamma_-/2]$.

\subsection{Switching probabilities}

In the secular approximation,  for preparation
at $t=0$ in $| +- \rangle $ and  for $k_B T \ll \Omega$,
the probabilities \eref{eq:psw1} and \eref{eq:psw2} take the simpler form
\begin{eqnarray}
\label{switch1}
P_{\mathrm{SW}1}(t) = &-& \frac{1}{2} \, \cos \varphi \,
\left [ \rho_{11}(t) + \rho_{22}(t)  \right ]
+ {\rm Re}[\rho_{12}(t)] + \cos^2\left (\frac{\varphi}{2}\right )
\\
\label{switch2}
P_2(t) = &- & \frac{1}{2} \, \cos \varphi \,  
\left [ \rho_{11}(t) + \rho_{22}(t)  \right ]
- {\rm Re}[\rho_{12}(t)] + \cos^2\left (\frac{\varphi}{2}\right ) \, .
\end{eqnarray}
where $\rho_{11}(t)$ and $\rho_{22}(t)$ are given by eqs.~\eref{population1}, \eref{population2} 
and 
$$\rho_{12}(t)= \rho_{12}(0) \exp{\{- t/T_2^{SWAP} - i \tilde \omega_{12} t \} } \, .$$
Anti-correlation of the above probabilities directly follows from the coherence $\rho_{12}(t)$
entering with different signs in $P_{\mathrm{SW}1}(t)$ and in
$P_2(t)$. In order to check the efficiency of the gate 
we will therefore consider only the qubit 1 switching probability.
Neglecting the frequency shift of $\omega_{21}$ (see \ref{appendix:shifts}), 
$P_{\mathrm{SW}1}(t)$ can be approximated as
\begin{equation}
P_{\mathrm{SW}1}(t) \approx \frac{1}{2} \, [e^{- t/T_R} \, -\, \cos(\omega_{21} t) 
\,e^{-  t/T_2^{SWAP}} ] \, + \, \cos^2\left (\frac{\varphi}{2}\right ) \, .
\end{equation}
Here we explicitly see that, in order to 
perform the $\sqrt{{\rm i-SWAP}}$ 
operation, it is necessary that $T_R \, , \,  T_2^{SWAP} \gg 1/\omega_c$. Efficient
entanglement generation is guaranteed when $T_R \gg T_2^{SWAP}\gg 1/\omega_c$, i.e. when
transverse and longitudinal quantum noise are such that 
$S_{x_\alpha}(\Omega) \ll S_{z_\alpha}(\omega_c)$.
Under this condition, the efficiency of the gate is limited by $T_2^{SWAP}$, i. e. by
longitudinal noise in the computational basis, $S_{z_\alpha}(\omega_c)$, which is
responsible for the short-times behavior.
Decay towards the equilibrium value 
$P_{\mathrm{SW}1}(\infty) = \cos^2 (\varphi/2)= (1+ \cos \varphi)/2= (1- 1/\sqrt{1+
(\omega_c/2\Omega)^2})/2 \approx (\omega_c/4 \Omega)^2$
occurs in a time of the order of $T_R$, due to transverse noise in the computational basis.
\begin{table}
\caption{
\label{tab:noise} 
Noise characteristics deduced from single qubit data reported in ref.\cite{vion}. 
Since both qubits operate at $q_{\alpha,\rm{x}}=1/2$, they are presumably 
affected by similar polarization fluctuations, thus $\sigma_{x_\alpha}$ and $S_{x_\alpha}$
are the same as in \tref{tab:noise1}. Since the two qubits operate at different phase points,
their phase spectral characteristics differ.
From ref.\cite{vion} the power spectrum of phase fluctuations is
$S_{\delta}(\omega) \approx [ 5.57 \times 10^{-7}]/\omega \, + \, 3.7 \times 10^{-14} \mathrm{[s]}$
where the extrapolated crossover frequency is $\approx 2 \pi 10$~MHz. 
Since $\delta_2 \neq 0$  the process $\hat z_2 \propto \Delta \delta_2$ is Gaussian and 
characterized by  
$S_{z_2}(\omega) =  ( E_{J,2}^0  \sin \delta_2 )^2 (\tilde \Phi_{2,++} - \tilde \Phi_{2,--})^2 S_{\delta_2}(\omega)$
(where $\tilde \Phi_{2, \pm \pm}$ are defined in \ref{appendix-CPB}, \tref{tab:meanvalues}). 
Instead  $z_1 \propto (\Delta \delta_1)^2$ is non Gaussian and its spectrum is ohmic,
$S_{z_1}(\omega) = 2 S^2/(\pi \Omega^2) \, \omega \, 
\coth(\omega / (2 K_B T))$, where $S= 1.6 \times 10^8$s$^{-1}$.
In this table we fixed $\Omega \approx   10^{11}$rad/s, $\omega_c \approx 10^{-2} \Omega$
and $T= 40$~mK.
}
\begin{indented}
\item[]\begin{tabular}{ccc}
\br
 	& Qubit 1 & Qubit 2 \\
\mr
%\hline
$S_x^{1/f}$ 	&  $ \sigma_{x_1} \approx 2 \times 10^{-2} \,\Omega$      &  $ \sigma_{x_2} \approx 2 \times 10^{-2} \,\Omega$ \\
$S_z^{1/f}$  	&  $\sigma_{z_1} \approx 10^{-6} \, \Omega $  & $\sigma_{z_2} \approx 6 \times 10^{-4} \, \Omega$\\
\hline
$S_x^f$		& $S_{x_i}(\Omega) \approx 4 \times 10^6$s$^{-1}$   & $S_{x_i}(\Omega) \approx 4 \times 10^6$s$^{-1}$ \\
$S_z^f$		 & $S_{z_1}(\omega_c) \approx 10^4$ s$^{-1}$
& $S_{z_2}(\omega_c)  \approx 5 \times 10^7$s$^{-1}$\\
%\hline
\br
\end{tabular}
\end{indented}
\end{table} 

\begin{figure}[t!]
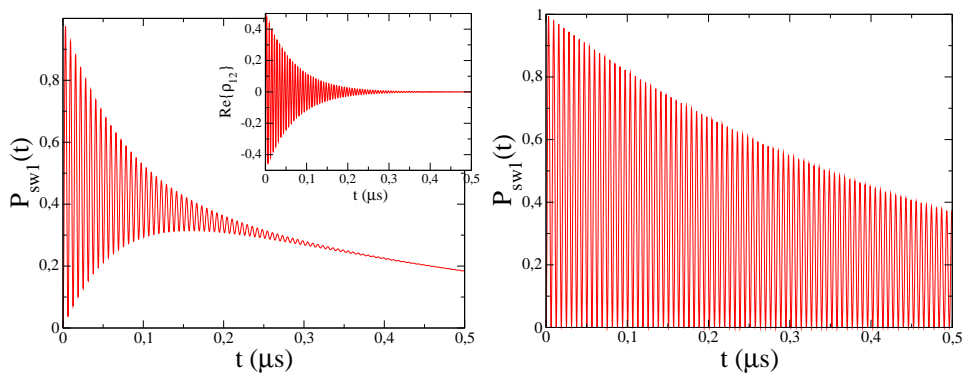

\centering
\includegraphics[width=0.4\textwidth]{figure3a.eps}
\includegraphics[width=0.4\textwidth]{figure3b.eps}
\caption{Left panel: Switching probability of qubit 1 for the noise levels $S_x^f$ and 
$S_z^f$ reported in \tref{tab:noise}. 
The long-time decay is due to charge noise entering the populations of levels $1$ and $2$. Inset: 
the real part of the SWAP coherence which is responsible for the short-time behavior of
the switching probability due to phase noise on qubit $2$. 
Right panel: Switching probability of qubit 1 in the presence of charge noise
with white spectrum, 
$S_{x_i}(\omega) \approx 4 \times 10^6$s$^{-1}$.
}
\label{fig1}
\end{figure}

\noindent \underline{Capacitively coupled CPB-based qubits}: The relevant rates 
\eref{eq:ratesgeneral} can be estimated  in this specific case and are reported 
in \ref{appendix-CPB}.
For the charge-phase two-ports architecture, control is via the gate voltage,
$q_x = C_g V_g/(2e)$ and magnetic flux dependent phase, $\delta$, entering the Josepshon
energy. The single qubit optimal point is at $q_x=1/2$, $\delta=0$. 
The resonant condition between the two qubits with a capacitive coupling
is achieved by displacing  one of the two qubits from the "double" optimal point. 
In order to limit the sensitivity to charge noise, resonance is achieved tuning 
$\delta_2 \approx 0.45$. 
The noise characteristics are reported in \tref{tab:noise}, where we note that
$S_{x_\alpha}(\Omega) \ll S_{z_\alpha}(\omega_c)$.
Because of the operating conditions, absorption (and emission) rates due to phase noise on qubit 2 
dominate over rates due both to charge noise and to  phase noise on qubit 1.
Relaxation of the populations takes place on a scale
$1/T_R = \Gamma_+ \approx  \Gamma_{10} \approx \Gamma_{20} \approx S_{x_i}(\Omega)/2$
 due to transverse (charge) noise on both qubits, whereas the SWAP coherence decay rate
 is dominated by longitudinal (phase) noise on qubit 2,
$1/T_2^{SWAP}= \widetilde \Gamma_{12} \approx \frac{1}{2} (\Gamma_{12} + \Gamma_{21}) \approx \Gamma_{21} 
\approx S_{z_2}(\omega_c)/4$.
The efficiency of the $\sqrt{{\rm i-SWAP}}$ gate is  mainly limited by phase noise on qubit 2,
which gives $T_2^{SWAP} \approx 1/\Gamma_{21} \approx 100$~ns. 
We note that, under these conditions, $T_2^{SWAP}$ is comparable with the decoherence time 
of qubit 2,  $T_2=1/[S_{x_2}(\Omega)/4 + S_{z2}(0)/2] \approx 2/S_{z2}(0) \approx
2/S_{z2}(\omega_c)$.
The  phase-dominated behavior is followed by a slower charge-dominated decay towards the 
equilibrium value 
$P_{\mathrm{SW}1}(\infty) \approx (\omega_c/4 \Omega)^2$. 
These features are illustrated in
\fref{fig1} (left) where for comparison the real part of the SWAP coherence is shown in the inset.
We note that the contribution of  high-frequency charge noise to the efficiency of the 
$\sqrt{{\rm i-SWAP}}$ gate is relatively small. Indeed if only polarization fluctuations 
were present, we could approximate
 \begin{equation}
P_{\mathrm{SW}1}(t) \approx \frac{1}{2} \, [1 \, -\, \cos(\omega_{21}  t) ] 
\, e^{- t/T_2^{SWAP}}\, + \, \cos^2\left (\frac{\varphi}{2}\right )
\end{equation}
and $T_2^{SWAP} \approx 1/ \Gamma_{10}\approx 0.5$~$\mu$s. This situation is illustrated in 
\fref{fig1} (right).
Note that for the estimated noise figures, the condition  for the secular approximation,
$\omega_{21} = \omega_c \gg \widetilde \Gamma_{12}$, is satisfied.

\subsection{Concurrence}

For the considered initial condition, the RDM takes the "X-form"~\cite{Xform} and 
the concurrence is given by
\begin{eqnarray}
C(t) \approx  {\rm max} &\Big [& 0, 
\sqrt{ ( \rho_{11}(t) - \rho_{22}(t) )^2 +  (2{\rm Im }[\rho_{12}(t)])^2 }-|\sin \varphi| \rho_{00}(t),
\nonumber \\
&& |\sin \varphi| \rho_{00}(t) -
\sqrt{ ( \rho_{11}(t) + \rho_{22}(t) )^2 -  (2{\rm Re }[\rho_{12}(t)])^2 } \Big ] \, .
\label{concurrence-general}
\end{eqnarray} 
At times shorter than the global relaxation time, $T_R$, the ground state is almost
unpopulated, $\rho_{00}(t) \approx 0$, and $C(t)$ is given by the second term in \eref{concurrence-general}
\begin{equation}
\label{conc-quantum}
C(t) \approx \left[ \frac{\Gamma_+}{\Gamma_-} \, e^{-2t/T_R} + \sin^2( \omega_{12}t) \, 
\,e^{-2t/T_2^{SWAP}} \right]^{1/2} - |\sin \varphi | \, (1- e^{-t/T_R}) \, ,
\end{equation}
for $t \ll T_R$, the concurrence is approximately
given  by the SWAP coherence 
$C(t) \approx 2 |{\rm Im}\{\rho_{12}(t)\}| = \sin( \omega_{12}t) \, e^{-t/T_2^{SWAP}}$.
Like the switching probabilities, the concurrence evolves with the SWAP coherence 
decay time, $T_2^{SWAP}$, due to originally longitudinal noise.

\begin{figure}[t!]
\centering
\includegraphics[width=0.5\textwidth]{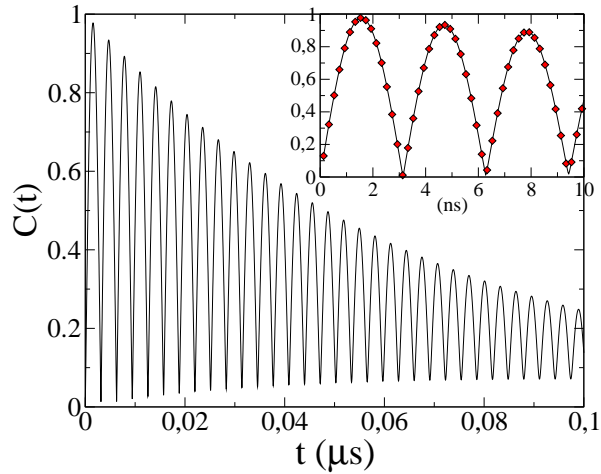}
\caption{Concurrence given by \eref{conc-quantum} for $\omega_c/\Omega= 0.01$ and
for quantum noise values  $S_x^f$ and 
$S_z^f$ reported in \tref{tab:noise}.
Inset: At short times  
$C(t) \approx 2 |{\rm Im}\{\rho_{12}(t)\}|$ (diamonds).
}
\label{fig4}
\end{figure}

\noindent \underline{ Capacitively coupled CPB-based qubits}:
The concurrence  for the charge-phase 2-qubit gate is illustrated in  \fref{fig4}. 
We note that the long-time behavior is instead due to populations relaxation to the ground
state and $C(t)$ is given by  the third term in \eref{concurrence-general}
\begin{eqnarray}
C(t) &\approx& 
|\sin \varphi | \, (1- e^{-t/T_R}) - 
\left[  \, e^{-2t/T_R} - \cos^2( \omega_{12}t) \, \,e^{-2t/T_2^{SWAP}} \right]^{1/2}  
\nonumber \\
&\approx& - |\sin \varphi| \rho_{00}(t) \to |\sin \varphi |
\end{eqnarray}
The finite asymptotic value $C(t \to \infty) \approx |\sin \varphi|$, reflects  
the entangled thermalized state. Because of the interaction between the two qubits
the phenomenon of entanglement sudden death does not take place~\cite{Xform}.  

\section{Adiabatic noise}
\label{sec:adiab}

Let's consider now the effect of low frequency fluctuations, replacing in \eref{int-plit}
$\hat{E}_{\alpha} \, \to \, \hat E_\alpha^{\mathrm A} \equiv   E_{\alpha}(t)$.
In the adiabatic and longitudinal approximation~\cite{PRL05,PhysScript09} populations do not
evolve and the system dynamics 
is related to instantaneous eigenvalues, $\omega_i(\vec E(t))$, which
depend on the noise realization, $\vec E(t)$. They enter the coherences
in the eigenbasis of ${\cal H}_0$  in the form
\begin{equation}
\rho_{ij}(t) = \rho_{ij}(0) \int \, \mathcal{D}[\vec E(s)] \, P[\vec E(s)] \, e^{-i \int_0^t ds \,
\omega_{ij}(\vec E(s))} \,
\label{path-int}
\end{equation}  
where the probability of the realization $\vec E(s)$, $P[\vec E(s)]$, also depends on the 
measurement protocol. 
A standard approximation of the path-integral \eref{path-int} consists in
replacing $E_{\alpha}(t)$ with statistically distributed values 
$E_{\alpha}(0) \equiv E_{\alpha}$ at each repetition of the measurement protocol. 
The "static-path approximation" (SPA)~\cite{PRL05} or "static-noise" approximation~\cite{ithier} 
gives the leading order effect of low-frequency fluctuations in repeated measurements. 
In the SPA, the level splittings $ \omega_{ij}(\vec E)$ are random variables, with 
standard deviation $\Sigma_{ij}=\sqrt{\langle\delta \omega^2_{ij}\rangle-
\langle\delta \omega_{ij}\rangle^2}$, where $\delta \omega_{ij} = \omega_{ij}(\vec E)-
\omega_{ij}$. 
The coherences \eref{path-int} reduce to ordinary integrals
\begin{equation}
\rho_{ij}(t) \approx \rho_{ij}(0) \int d \vec E  
P(\vec E) \; e^{- i \omega_{ij}(\vec E) t} \, \equiv 
\rho_{ij}(0) \, \langle  e^{- i \omega_{ij}(\vec E) t}  \rangle \, ,
\label{SPA-general}
\end{equation}
where the probability density,  in relevant cases, can be taken of Gaussian
form, $P(\vec E) \equiv \Pi_\alpha P(E_{\alpha})$ with
$P(E_{\alpha}) = \exp[- E_\alpha^2 /2 \sigma_{E_\alpha}^2]/\sqrt{2 \pi} \sigma_{E_\alpha}$~\cite{PRL05}. 
The splittings  $ \omega_{ij}(\vec E)$ come  both from "effective longitudinal"  and
from "effective transverse" terms in (\ref{HI-proj}). This is analogous to a single qubit, where
longitudinal noise gives the leading order linear terms and transverse noise is responsible
for second order terms which dominate at the optimal point, where the first order
longitudinal contributions vanish~\cite{ithier,Makhlin05}. Here, the Z-splitting $\omega_{03}(\vec E)$
has a linear contribution due to the effective longitudinal noise  
$(\hat{z}_{1} + \hat{z}_{2} )  \, \cos \varphi  \, ( |0\rangle \langle 0| - | 3\rangle \langle 3|) $
in \eref{H-trunc-Z}.  The  SWAP splitting $\omega_{21}(\vec E)$ instead,
in  the absence of leading-order effective-longitudinal intra-doublet terms in \eref{H-trunc-SWAP},
comes from higher order contributions due to "effective transverse" noise.
Evaluating them requires considering the complete Hilbert space of the coupled qubit system
(inter-doublet processes included). 
The systematic approach to obtain these contributions consists in
treating in perturbation theory effective transverse terms in $\mathcal{H}_\mathrm{I}$,
where, in the adiabatic approximation, $\hat x_\alpha$ and  $\hat z_\alpha$ are 
replaced by  classical stochastic fields $x_{\alpha}$ and 
$z_{\alpha}$.
We obtain
\begin{eqnarray}
&& \!\!\!\!\!\!\!\!\! \omega_{21}(x_{1},x_{2},z_{1},z_{2})
\approx \omega_c -\frac{\omega_c}{2 \Omega^2}  ( x_{1}^2 + x_{2}^2 ) 
+ \frac{1}{2\omega_c}(z_1 - z_2)^2 
	\nonumber \\
 &+& \!\! \frac{\omega_c}{2 \Omega^3}  (x_1^2 + x_2^2) (z_1 + z_2) +  
        \frac{1}{2\omega_c\Omega}  (x_1^2  -  x_2^2)(z_1 - z_2)
 \label{eq:splittingSWAP} \\
 &+&  \!\!   \frac{\omega_c}{8 \Omega^4} (1+  \frac{\omega_c^2}{\Omega^2})
( x_1^4 + 6 x_1^2 x_2^2  +  x_2^4) 
+ \frac{1}{8\omega_c \Omega^2}( x_1^2  -  x_2^2 )^2  \, , \nonumber \\
&& \nonumber \\
&& \!\!\!\!\!\!\!\!\!  \omega_{03}(x_{1},x_{2},z_{1}, z_{2}) \approx 
2 \sqrt{\frac{\omega_c^2}{4}+ \Omega^2} \, 
- \, \cos \varphi \, (z_{1}+ z_{2})
\nonumber \\
&+& 
\frac{1}{2 \Omega} \left ( \frac{\omega_c}{\Omega} \sin \varphi - \cos \varphi \right )
\,(x_1^2  +  x_2^2)
+  \frac{\omega_c}{ 8\Omega^2} \sin \varphi \, (z_1+ z_2)^2 \, ,
\label{eq:omega03-order2}
\end{eqnarray}
where $\tan \varphi = - \omega_c/(2 \Omega)$.
The SWAP-splitting has been evaluated  up to $4^{th}$ order~\cite{NASA}, whereas 
the Z splitting expansion is considered up to second order since the 
fourth-order terms are much smaller, scaling with $\omega_c^3$.
The Z-splitting consists of a linear term due to the effective longitudinal noise,
which dominates with respect to the quadratic term due to effective transverse
intra-subspace noise (scaling with $\sin \varphi$). Transverse inter-doublet noise
gives an additional second order contribution.

Performing the average \eref{SPA-general} 
(including in the $3^{rd}$ and $4^{th}$ order terms  in \eref{eq:splittingSWAP} 
only the contributions $\propto \omega_c^{-1}$)
we obtain the SWAP coherence in the SPA
\begin{equation}
\rho_{12}^{SPA}(t) = \rho_{12}(0)  \,
\frac{ \Omega }{2 \sigma_x^2} \sqrt{\frac{2 i
\omega_c}{\pi  t}} \, 
e^{i \omega_c t+ h(t)} \, K_0[ h(t)] 
\label{eq:SPA-SWAP}
\end{equation}
where $h(t)= (\sigma_{z_1}^2 +\sigma_{z_2}^2 + i \omega_c/t) \, 
(\Omega^2/\sigma_x^2 +i \omega_c t)^2/(4 \Omega^2)$ and
 $K_0[h]$  is the K-Bessel function of order zero~\cite{abramowitz}.
In the absence of mixed terms in the expansion
\eref{eq:omega03-order2}, different contributions to the Z-coherence
factorize and lead to
\begin{equation}
\!\!\!\!\!\!\!\!\!\!\!\!\!\rho_{03}^{SPA}(t) = \rho_{03}(0)  
e^{ 2 i\sqrt{\frac{\omega_c^2}{4}+ \Omega^2} t}\; 
\frac{e^{- \frac{|\cos \varphi|}{2}  \, (\sigma_{z_1}^2 + \sigma_{z_2}^2) t^2}}
{1- 
\frac{i}{\Omega} \left ( \frac{\omega_c}{\Omega} \sin \varphi - \cos \varphi \right )
 \, \sigma_x^2 t} \, .
\label{eq:SPA-Z}
\end{equation}
Here we assumed the same variance for the transverse noise components, 
$\sigma_{x_1}= \sigma_{x_2}\equiv \sigma_x$. 
Instead we maintained $\sigma_{z_1}$ and $\sigma_{z_2}$ distinct, considering that the 
two qubits may operate at different working points (\ref{appendix-CPB}). 
In \eref{eq:SPA-Z} the exponential factor comes from linear terms in \eref{eq:omega03-order2} due to
effective longitudinal noise in the Z-subspace. 
The algebraic decay instead comes from quadratic terms in the expansion of
$\omega_{03}$, due to effective transverse inter-doublet processes.  \\
We remark that the applicability of the Gaussian approximation to the fields $E_\alpha$ 
depends on the relation between each $E_\alpha$ and the fluctuations
of the system's physical parameters.
For instance, for a charge-phase $\sqrt{{\rm i-SWAP}}$ gate the resonant condition occurs for 
$\delta_1=0$, $\delta_2 \neq 0$ (\ref{appendix-CPB}).
Thus it is $z_1 \propto  E_{1,\mathrm{J}}^0 (\Delta  \delta_1)^2$, where
$\delta_1$ (not $z_1$) is reasonably assumed Gaussian distributed. 
This would result in a modification of the terms $\propto \sigma_{z_1}$
in Eqs. \eref{eq:SPA-SWAP} and \eref{eq:SPA-Z}.
For instance, the integral over $z_1$  in $\rho_{03}^{SPA}$  would reduce to
$\{1-i  \sqrt{2} \cos \varphi \, \sigma_{z_1} t\}^{-1/2}$,
instead of $\exp{\{- |\cos \varphi| \sigma_{z_1}^2 t^2/2 \}}$.
The quantitative effect in $\rho_{03}^{SPA}(t)$ (and similarly in $\rho_{12}^{SPA}(t)$) is however 
negligible, because of the smallness of longitudinal noise at the optimal point, 
$\sigma_{z_1} \ll \sigma_{z_2}$ (see \tref{tab:noise}).

Validity regime of the SPA: The adiabatic approximation is tenable for times
shorter than the relaxation times, for this problem this condition requires that
$t \ll T^{SWAP}_1$. 
The static approximation is exact for times smaller that $1/\gamma_M$ (in case of a sharp high
frequency cut-off). 
We verified that it is a good approximation also for times $ t > 1/\gamma_M $ if $\gamma_M < \Omega$
and the $1/f$ spectrum is originated from an ensemble of bistable fluctuators with switching rates
$\in [\gamma_m, \gamma_M]$, leading to a $1/f^2$ decay above $\gamma_M$
(numerical simulations and analytic first correction to the SPA  \cite{PRL05}).  

\subsection{Minimization of defocusing: optimal coupling}

Exploiting the band structure of coupled nano-devices it is possible to reduce
the influence of $1/f$ fluctuations~\cite{PRB10}. The basic idea of "optimal tuning" is to fix
control parameters to values which minimize the variance of the splittings 
$ \omega_{ij}(\vec E)$, $\Sigma_{ij}^2$.
This naturally results in a enhancement of the decay time of the corresponding coherence
due to inhomogeneous broadening, i. e. of $\rho_{ij}^{SPA}(t)$. This is simply understood
considering the short times expansion of 
$\langle  \exp{\{- i \delta \omega_{ij}(\vec E) t\}}  \rangle  $ 
\begin{equation}
\langle  e^{- i \delta \omega_{ij}(\vec E) t}  \rangle \approx 1 -i \langle \delta \omega_{ij}(\vec E) \rangle t 
-\frac{1}{2} \langle \delta \omega_{ij}(\vec E)^2 \rangle t^2 
\end{equation}
the short-times decay of the coherence in the SPA is therefore given by
\begin{equation}
|\langle  e^{- i \delta \omega_{ij}(\vec E) t}  \rangle | \approx \sqrt{1- (\Sigma_{ij} t)^2} \, ,
\end{equation}
resulting in reduced defocusing for minimal variance $\Sigma_{ij}$.
For a single-qubit gate, the "optimal tuning" idea immediately leads to the
well-known "magic point". 
In fact, if $ \omega_{ij}(\vec{E})$  is monotonic 
in a region $|E_\alpha| \leq 3 \sigma_{E_\alpha}$, we can approximate
$\Sigma_{ij}^2 \approx \sum_\alpha \left [\frac{\partial \omega_{ij}}{\partial
E_\alpha}|_{E_\alpha=0}\right ]^2 \sigma_{E_\alpha}^2$, thus
the variance attains a minimum for vanishing differential dispersion.
For the charge-phase two-port architecture, control is 
via gate voltage, 
$q_{x}$, and magnetic-flux dependent phase $\delta$,
thus $E_\alpha$ corresponds to the fluctuations $\Delta q_{x}, \Delta E_J $ and
the optimal point, $q_x=1/2 \, , \delta=0$, 
is at the a saddle point of the energy bands~\cite{vion}.
When bands are non-monotonic in the control parameters, 
minimization of defocusing necessarily requires their 
tuning  to values depending on the noise variances.
For a multi-qubit gate, the optimal choice has to be done considering the 
most relevant coherence for the considered operation. 
Here we show how this program applies to the coherence in the SWAP subspace and
partly to $\rho_{03}^{SPA}(t)$.
\begin{figure}
\begin{center}
\includegraphics[width=7cm]{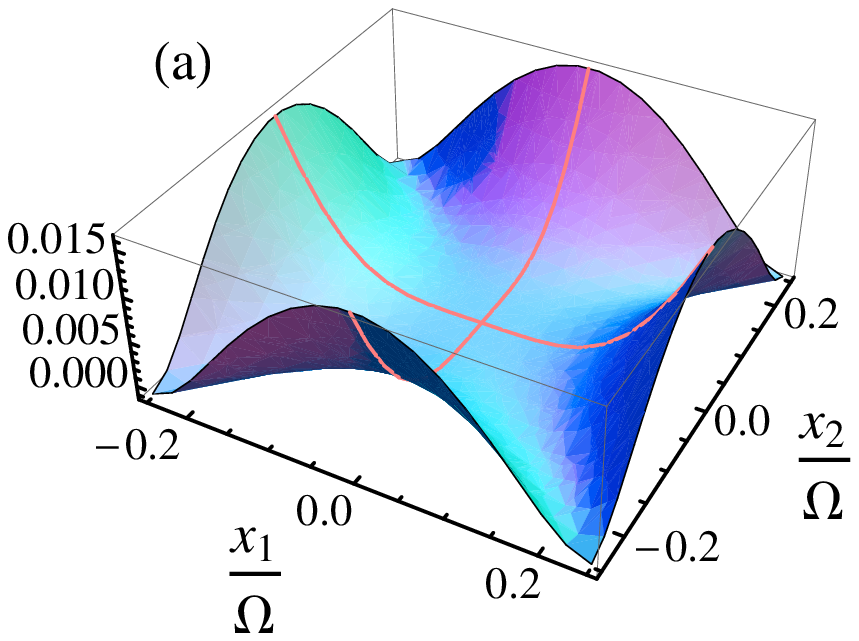}
\includegraphics[width=7.5cm]{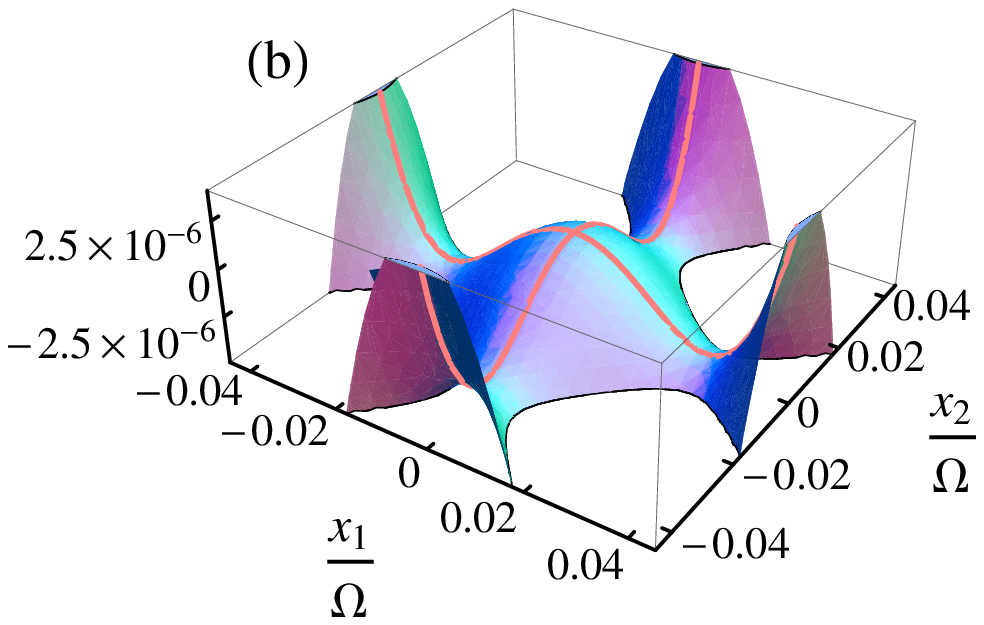}
\includegraphics[width=7cm]{figure5c}
\includegraphics[width=7cm]{figure5d}
\end{center}
\caption{ Dispersion of the SWAP and Z splittings for $\omega_c/\Omega=0.02$. 
Panels (a) and (b): $\delta \omega_{21}/\Omega$ from numerical diagonalization of 
$\mathcal{H}_0 + \mathcal{H}_\mathrm{I}$. In panel (b) a zoom around the origin 
highlights the interplay of $2^{nd}$ and $4^{th}$ order terms of the expansion
\eref{eq:splittingSWAP}, the barrier height is $\propto \omega_c^3$.
%Panels (c) and (d): Comparative behavior of dispersions in the two subspaces.
Panel (c): SWAP exact splitting  (blue), expansion (\ref{eq:splittingSWAP}) for 
$x_2=0$, $z_i=0$  (dashed), $2^{nd}$ order expansion (dash-dotted green), $Z$ splitting (red)
from \eref{eq:omega03-order2}
and single qubit dispersion  (diamonds).
Panel (d): Longitudinal dispersions in the SWAP (blue) and Z (red) subspaces.
}\label{fig-bands}
\end{figure}

A key feature is that the SWAP splitting, $\omega_{21}$ in Eq. \eref{eq:splittingSWAP}
is non-monotonic in the small coupling  $\omega_c \ll \Omega$. 
This is due to the fact that effective transverse fluctuations originate
both from longitudinal and transverse noise, and give rise respectively to
intra-SWAP and inter-doublet transitions, see \eref{H-trunc-SWAP}, \eref{H-inter}.
For instance, second order corrections to $\omega_1$
from effective transverse intra-doublet processes are
$\omega_c^{-1} \propto |\langle 1 | \mathcal{H}_\mathrm{I} |2 \rangle |^2/(\omega_1-\omega_2)$,
from effective transverse 
inter-doublet transitions are 
$\sum_{i \neq 1,2}  | \langle 1 | \mathcal{H}_\mathrm{I} |i \rangle|^2 /(\omega_1- \omega_i) \propto
\omega_c$. 

Non-monotonicity in $\omega_c$ results in a competition between 
$2^{nd}$ and $4^{th}$ order $x_{\alpha}$-terms in \eref{eq:splittingSWAP} and in
non-monotonic band structure,  \fref{fig-bands} (panels (a) and (b)). 
Because of this subtle feature, identification of the best operating condition  
necessarily requires consideration of the noise characteristics.
Indeed the {\em optimal coupling}  which minimizes the SWAP variance 
\begin{equation}
\Sigma_{21}^2 \approx \frac{1}{\omega_c^2} 
 \left \{  \left (\frac{\sigma_x}{\Omega} \right )^4 
\left [ (\sigma_x^2 - \omega_c^2)^2 + \sigma_x^4 
+ \sigma_{z_2}^2 \Omega^2 \right ] + \frac{\sigma_{z_2}^4}{2}   \right \} 
\label{eq:variance}
\end{equation}
is given by
\begin{equation}
\widetilde \omega_c = \left \{ 2 \sigma_x^4 + \sigma_{z_2}^2 \Omega^2 + \frac{1}{2} \left ( \frac{\sigma_{z_2}
\Omega}{\sigma_x} \right )^4 \right \}^{1/4} \, ,
\label{optimal}
\end{equation}
where we assumed $\sigma_{z_1} \ll \sigma_{z_2}$.
The effectiveness of the optimal coupling choice has been discussed in details in ref. \cite{PRB10}.
The advantage of operating at optimal coupling with respect to a generic $\omega_c$
can be parametrized by the error of the gate at time  
$t_{e}= \pi /2 \omega_c$
when system should be in the entangled state 
$| \psi_e \rangle = [\vert +- \rangle - \rmi \vert -+ \rangle]/\sqrt 2$,
$\varepsilon = 1 - \langle \psi_e | \rho(t_e) | \psi_e \rangle$. 
As shown in \tref{error} for two CPB-qubits, in the presence of moderate
amplitude transverse (charge) noise, at the optimal coupling the error can be reduced even 
one order of magnitude with respect to a generic coupling.  
\begin{table}
\caption{
\label{error} 
Error at the first $\sqrt{{\rm i-SWAP}}$ operation at time $t_{e}$ for various couplings
$\omega_c/\Omega$ (numerical simulation of the coupled dynamics). 
The error $\varepsilon_a$  refers to a typical amplitude $\sigma_x = 0.02 \Omega$;
the error $\varepsilon_b$  refers instead to a moderate amplitude $\sigma_x = 0.04 \Omega$. 
Longitudinal noise is here $\sigma_{z_2}= 10^{-3} \Omega$. The optimal coupling,\eref{optimal},
is in both cases at $\widetilde \omega_c \approx 0.05 \Omega$. 
The error $\varepsilon_a$  is reduced by increasing $\omega_c$ because of the comparatively large effect 
of phase noise. In the case of $\varepsilon_b$, the most detrimental effect is from charge 
noise and the error is minimum at $\widetilde \omega_c$.}
\begin{indented}
\item[]
\begin{tabular}{ccc}
\br
 $\omega_c/\Omega$ & $\varepsilon_a$ & $\varepsilon_b$ \\
\hline
 $0.01$	&  $3 \cdot 10^{-3}$    &  $10^{-2}$   \\
 $0.02$ 	&  $1.5 \cdot 10^{-3}$  &  $5 \cdot 10^{-3}$  \\
 $0.04$	&  $6 \cdot 10^{-4}$    &  $2 \cdot 10^{-3}$ \\
 $\it{0.05}$ & $\it{  8 \cdot 10^{-4}}$ &   $\it{8 \cdot 10^{-4}}$ \\
 $0.06$	&  $3 \cdot 10^{-4}$     & $10^{-3}$\\
 $0.08$ &  $2 \cdot 10^{-4}$     & $9 \cdot 10^{-4}$\\
%\hline
\br
\end{tabular}
\end{indented}
\end{table} 

\begin{figure}[t!]
\centering
\includegraphics[width=0.5\textwidth]{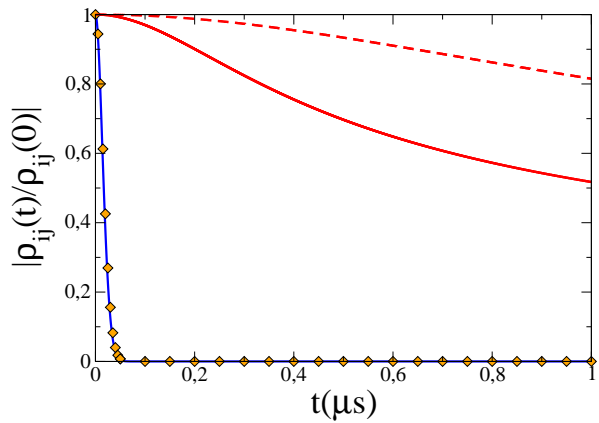}
\caption{Coherence in the SWAP-subspace (continuous red), in the Z-subspace (blue) 
and the single qubit-coherence (orange diamonds) for generic coupling $\omega_c= 0.01 \Omega$. 
The dashed red line is the coherence in the SWAP-subspace for optimal coupling, 
$\tilde \omega_c\approx 0.03 \Omega$. 
The Z coherence does not appreciably change by changing $\omega_c$ and decays
similarly to a single qubit at the double optimal point.
Noise characteristics are reported in \tref{tab:noise}.
}
\label{coherences}
\end{figure}

The splitting in the Z-subspace, on the contrary, is monotonic both in $x_{\alpha}$ and
in $z_{\alpha}$, similarly to a single qubit operating at $q_{\alpha, \mathrm x}=1/2$ and
at the two different $\delta_\alpha$, \fref{fig-bands}, panels (c) and (d). 
The most relevant contribution to Z-variance is given by the effective longitudinal noise
\begin{equation}
\Sigma_{03}^2 \approx \cos^2 \varphi \, \sigma_{z_2}^2 \, ,
\label{Z-variance}
\end{equation} 
which can be reduced only by increasing the
qubit's coupling strength $\omega_c$, which is however limited by single qubit
splittings $\Omega$. Therefore, no special optimal point exists if the two-qubit
operation involves the dynamics inside the Z-subspace.
A comparison of the evolution of the coherences in the two subspaces and of the
single qubit-coherence is shown in \fref{coherences} for two CPB-based qubits. 
In order to keep the advantage of optimal tuning in the SWAP-subspace, the two-qubit
operation should not involve the Z-subspace. 

Finally, we remark that the above results only follow considering the 4-level spectrum 
of the coupled devices. 
In fact, because of mixing between subspaces, limiting the analysis to noise
projections inside the subspaces would miss a relevant part of the defocusing processes
coming from originally transverse fluctuations $\propto x_{\alpha}$, see 
eqs.~\eref{H0-proj}, \eref{HI-proj}.
This is clear if we consider the evolution of the SWAP coherences truncating the system Hilbert 
space to the SWAP subspace. Since the reduced
Hamiltonian is ~\eref{H-trunc-SWAP},  
effective transverse intra-doublet noise would lead to quadratic corrections to the
SWAP-splitting \footnote{This is analogous to a single qubit in the presence of adiabatic
transverse noise, which is equivalent to longitudinal quadratic noise~\cite{Makhlin05}.}.
In fact, treating in perturbation theory  $z_{\alpha}$  up to second order,
we would get $\omega_{21}(x_{1},x_{2},z_{1},z_{2}) \approx  \omega_c + \frac{1}{2\omega_c}(z_{1}-z_{2})^2 $
and the SWAP coherence would decay algebraically, as it is typically originated from "effective
transverse" low frequency noise
\begin{equation}
\rho_{12}(t) \propto \frac{1}{\sqrt{1- i (\sigma_{z_1}^2 + \sigma_{z_2}^2) t / \omega_c}} \, .
\end{equation}
Under this approximation,  reduction of defocusing could only be achieved by increasing the 
coupling strength.
Similarly, reducing the analysis to the Z-subspace, cfr eq.~\eref{H-trunc-Z},
would miss the quadratic dependence of the Z-splitting due to transverse noise.

\section{Time scales of the $\sqrt{{\rm i-SWAP}}$ operation: interplay of quantum and adiabatic noise}

In the multi-stage elimination approach, where quantum noise is traced out first
and adiabatic noise is retained as a classical stochastic drive, we have to replace in 
eqs.~\eref{population1} - \eref{population0} and in 
$\rho_{12}(t)= \rho_{12}(0) \exp{\{-  t/T_2^{SWAP} - i \tilde \omega_{12} t \}}$, 
$\omega_{ij}$ with $\omega_{ij}(\vec E(t))$ and perform the path-integral 
over the realizations of $\vec E(t)$. Note that the dependence on adiabatic noise enters
also the rates,  which should be averaged. If the dependence of the power spectra 
on the splittings is sufficiently smooth, it can be neglected and the effect of low 
frequency noise reduces to averaging phase factors in the coherences~\footnote{In the 
multistage approach, the classical variables $x_{\alpha}$ and  $z_{\alpha}$ enter parametrically also
in the frequency shifts resulting from the solution of the Master Equation. 
Here we neglect this dependence, see \ref{appendix:shifts}.}.  
Thus, provided that the dependence on 
$\vec E$ of $\widetilde \Gamma_{21}$ and $\widetilde  \Gamma_{30}$ can be neglected, the effects 
of quantum and adiabatic noise can be treated independently
and lead to
\begin{eqnarray}
\rho_{12}(t) &\approx & \rho_{12}^{SPA}(t) \, e^{- t/T_2^{SWAP} }
\label{eq:ro12} \\
\rho_{03}(t) &\approx & \rho_{03}^{SPA}(t) \, e^{-\widetilde  \Gamma_{30} t } \, .
\label{eq:ro03}
\end{eqnarray}
This result is confirmed by numerical solution of the stochastic Schr\"odinger
equation for classical fluctuations leading to dynamical $1/f$ noise ($\gamma_m= 2 \pi \times 1$~Hz
and $\gamma_M= 2 \pi \times (10^6 - 10^8)$~Hz).
The evolution of populations is only due to quantum noise (in the adiabatic approximation they 
do not evolve),  thus they are given by
Eqs. \eref{population1} - \eref{population0} and the switching probabilities are given by 
Eqs. \eref{switch1}, (\ref{switch2}) where $\rho_{ij}(t)$ are replaced by \eref{eq:ro12} and \eref{eq:ro03}. 
These matrix elements enter also the concurrence which reads,
for times $t \ll T_R$,
$C(t) \approx \sqrt{ ( \rho_{11}(t) - \rho_{22}(t) )^2 +  ({\rm Im }[2 \rho_{12}(t)])^2 } 
- |\sin \varphi| \rho_{00}(t)$.

The relevant time scales characterizing the efficiency
of the $\sqrt{{\rm i-SWAP}}$ operation depend both on the SWAP
coherence and on the populations of the first three levels. Due to the
interplay of adiabatic and quantum noise, the time dependence is not a superposition of
exponentials.
We can distinguish two time regions: a asymptotic long-time regime and a intermediate-to-short
time regime.
\begin{table}
\caption{
\label{table:decoherence} 
Relaxation and decoherence times of the $\sqrt{\rm{i-SWAP}}$ gate and
responsible physical processes
(long-to-intermediate time behavior of $C(t)$ and of the switching probabilities).
}
\begin{indented}
\item[]
\begin{tabular}{cc}
\br
Time scale  & Physical origin \\
\mr
Relaxation to the ground state
 & 
 Transverse noise 
 \\
  $T_R \approx 1/\Gamma_+$
 &
 at frequency $\Omega$
 \\
 \mr
 SWAP relaxation time 
 &  
 Longitudinal noise 
 \\
 $T_1^{SWAP}\approx 1/\Gamma_- $
 &
 at frequency $\omega_{21}$
 \\
 SWAP decoherence time 
  & 
   Longitudinal noise 
 \\ 
$T_2^{SWAP} \approx 2 T_1^{SWAP}$ 
 &
 at frequency $\omega_{21}$
 \\ 
SWAP dephasing time $T_2^{SWAP*}$ 
&
Low frequency noise 
\\
$\rho_{21}^{SPA}(T_2^{SWAP*})=e^{-1}$
&
(longitudinal AND transverse)
\\
\mr
SWAP total decoherence time &
if $T_R \gg T_1^{SWAP}$  i. e.
\\
$T_2^{SWAP} = [1/2 T_1^{SWAP} + 1/T_2^{SWAP*}]^{-1}$
& $S_x(\Omega) \ll S_z(\omega_{21})$\\
\br
\end{tabular}
\end{indented}
\end{table} 

The asymptotic behavior is entirely due to populations relaxation to the ground state, 
it is exponential and takes place with the "global" relaxation time $T_R\approx 1/ \Gamma_+$,
resulting from "effective transverse" inter-doublet processes,
whose order of magnitude is the spectrum of transverse fluctuations at frequencies of order $\Omega$.
In order to avoid leakage from the SWAP-subspace, any two-qubit operation  has to take place
on a time much shorter than $T_R$. This constraint also applies to the SWAP decoherence times.\\
\noindent The intermediate-to-short time behavior, $t \ll T_R$,  gives more relevant information on the gate
performance. In this time regime, relevant quantities are populations and coherences in the SWAP subspace.
We distinguish a intermediate time regime, characterized by 
 $T^{SWAP}_1 \approx 1/\Gamma_-$ and $T^{SWAP}_2 = 2 T^{SWAP}_1 \approx 1/\widetilde \Gamma_{12}$ 
due to {\em effective transverse} quantum noise inside this subspace, physically originated from longitudinal 
noise on each qubit. Adiabatic noise leads to additional decay of the SWAP
coherence,  $\rho_{12}(t) \approx \rho_{12}^{SPA}(t) \, e^{- t/T^{SWAP}_2 }$.
The resulting defocusing is analogous to a "pure dephasing" process,  we may name 
the typical time scale  $T^{SWAP *}_2$, defined as the time at which 
$|\rho_{21}^{SPA} (T^{SWAP *}_2)| = e^{-1}$.
The time $T^{SWAP *}_2$ is found by numerical inversion of \eref{eq:SPA-SWAP}.
The {\em intermediate-time} behavior of the two-qubit gate is characterized by 
$1/T^{SWAP}_1$ and $1/T^{SWAP}_2 = 1/(2T^{SWAP}_1)+ 1/T^{SWAP *}_2$, analogously to a two-state system.
These time scales and the responsible processes are summarized in \tref{table:decoherence}.
\begin{figure}[t!]
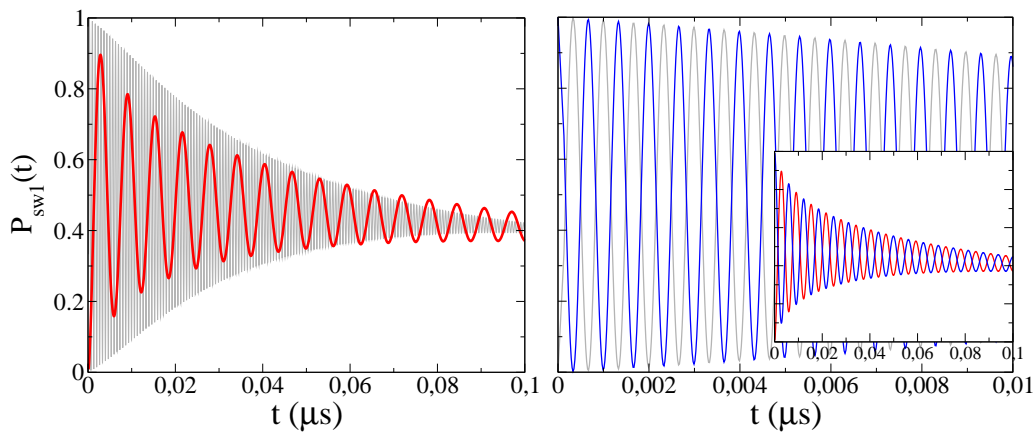

\centering
\includegraphics[width=0.45\textwidth]{figure7a}
\includegraphics[width=0.41\textwidth]{figure7b}
\caption{Qubit $1$ switching probability $P_{\mathrm{SW}1}(t)$ (red) and  probability $P_2(t)$ (blue)
to find  qubit $2$ in the initial state $\vert - \rangle$ in the presence of  $1/f$ 
and white noise for $\omega_c/\Omega= 0.01$ and for optimal coupling
$\tilde \omega_c = 0.08 \Omega$ (gray). 
To emphasize the robustness of the optimal coupling choice here
we consider $1/f$ charge noise of large amplitude, $\sigma_x = 8 \times 10^{-2} \Omega$. The
remaining noise figures are those reported in \tref{tab:noise}.
Left panel:  Plot of $P_{\mathrm{SW}1}(t)$ showing the 
exponential short-time behavior at $\tilde \omega_c$ and  the algebraic decay for generic
coupling.
Right panel: $P_{\mathrm{SW}1}(t)$ and $P_2$  anti-phase oscillations for
$\tilde \omega_c$ (main), $ \omega_c/\Omega= 0.01$ (inset).  
}
\label{switching}
\end{figure}

On the other side, in a quantum information perspective, it is important to estimate the time 
behavior (of concurrence or of switching probabilities) at times of the order of 
$t_e = \pi/2 \omega_c$, the first moment in which the entangled state is reached  by free evolution.
In order to generate entanglement within the SWAP subspace it is necessary that
$t_e \ll T_2^{SWAP}$.
It is therefore relevant to estimate the system behavior at short-times $t \ll T_2^{SWAP}$.
Both for the concurrence and the switching probability, the SWAP-coherence rules 
the short-time limit  which is approximately
\begin{equation}
\label{short-limit}
|\rho_{12}| \propto 1-  \frac{t}{ 2 T_1^{SWAP}} - \frac{1}{2} \, \Sigma_{21}^2 t^2 \, .
\end{equation}
The leading contribution, linear or quadratic, does not only depend on the noise amplitude but also on the operating point.
In fact $\omega_c$ enters the SWAP-splitting variance $\Sigma_{21}$ and, in principle, also
the SWAP decoherence time due to quantum noise $2 T_1^{SWAP}$. 
This is expected to be a smooth dependence. For the present considerations we assume a white 
spectrum in the relevant frequency range. 
For optimal coupling the short-time behavior is linear (since the
effect of low frequency noise is considerably reduced) whereas for a different coupling
strength the behavior is typically algebraic (see \fref{switching}).

\begin{figure}[ht!]
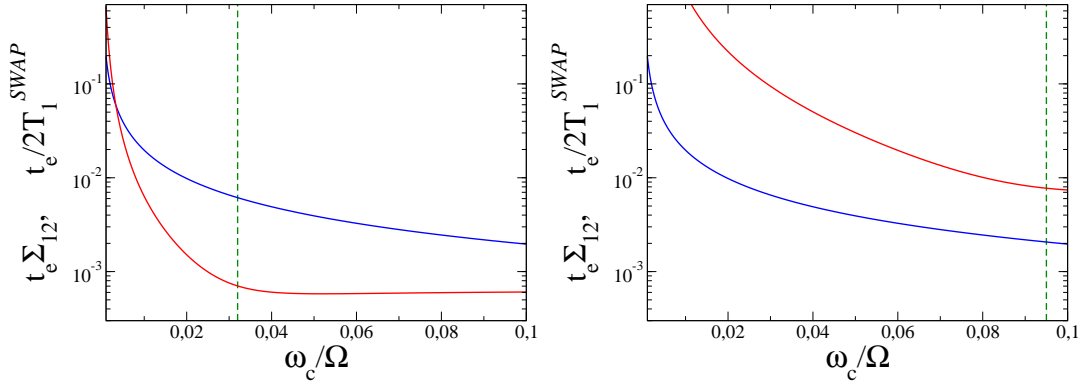

\centering
\includegraphics[width=0.45\textwidth]{figure8a}
\includegraphics[width=0.45\textwidth]{figure8b}
\caption{Plot of $\Sigma_{12}(\omega_c) t_e$ (red) and of $t_e/(2 T_1^{SWAP})$ (blue) as
a function of $\omega_c/\Omega$ (logarithmic scale) for $\sigma_x/\Omega= 0.02$ (left panel)
and $\sigma_x/\Omega= 0.08$ (right panel). The dashed line marks the optimal
coupling Eq.~\eref{optimal}.
}
\label{compare}
\end{figure}

The short-time expansion is valid at $t_e$ if  $t_e \ll T_2^{SWAP}$.
In \fref{compare} we consider two illustrative cases.
For the expected noise characteristics as reported in  \tref{tab:noise}, for couplings
$\Omega/100 < \omega_c < \Omega/10$
we have $\Sigma_{21} t_e  <  t_e/2 T_1^{SWAP} \ll 1$.
Thus the short-time expansion is valid up to $t_e$. Note that for the optimal
coupling  $1/2 T_1^{SWAP}$ is about one order of magnitude larger than 
$\Sigma_{21}$ (\fref{compare} left panel),
thus $|\rho_{12}| \propto 1-  t/2 T_1^{SWAP}$. For a larger amplitude of $1/f$
transverse noise, $\sigma_x \approx 0.08 \Omega$, we have
$ t_e/ 2 T_1^{SWAP} \leq \Sigma_{21} t_e \ll 1$ (\fref{compare} right panel). Also in this case the short-time expansion
holds up to $t_e$. In this case however even for optimal coupling the linear and quadratic terms
are comparable. The various possible short-times behaviors and validity regimes are
summarized in \tref{short}. 

\begin{table}[ht!]
\caption{
\label{short} 
Short times behavior $t \ll T_2^{SWAP}$ of $C(t)$ and of the switching probabilities
and validity conditions of the expansion up to $t_e$.
}

\begin{indented}
\item[]
\begin{tabular}{ccc}
\br
Short times  & Validity  & Expansion   \\
expansion  & conditions & up to $t_e$ if \\
\mr
$|\rho_{21}(t)| \approx 1 - \frac{t}{2 T_1^{SWAP}}$
 & 
Optimal coupling and $\Sigma_{21} \cdot 2  T_1^{SWAP} < 1$
&
$\frac{t_e}{2 T_1^{SWAP}} \ll 1$
 \\
 $|\rho_{21}(t)| \approx 1 - \frac{(\Sigma_{21} t)^2}{2}$
 &
 Generic coupling or  $\Sigma_{21} \cdot 2 T_1^{SWAP} > 1$
 &
 $ \frac{t_e }{1/\Sigma_{21}} \ll 1$
 \\
\br
\end{tabular}
\end{indented}
\end{table} 

\section{Conclusions}

In the present article we have identified the relevant decoherence times of 
an entangling operation due to the different decoherence channels originated 
from broad-band and non-monotonic noise affecting independently two qubits. 
Results depend on the interplay of noise at low and at high frequencies (with respect to both 
the single qubit splittings, $\Omega$, and the qubits coupling strength, 
$\omega_c$) and on the symmetries of the qubit-environment coupling Hamiltonian. 
In addition, "optimal" operating conditions for the universal two-qubit gate have been
identified.
Even if the relevant dynamics is within the SWAP subspace, the important time 
scales (both due to quantum and to adiabatic noise) can only be predicted 
considering the whole multilevel nature of the coupled systems.

In particular, relaxation processes from the SWAP subspace to the
ground state only depend on {\em transverse} noise at frequency $\Omega$. 
Decoherence processes internal to the SWAP subspace are instead originated from
{\em longitudinal} noise at frequency $\omega_c$.
This apparently counter-intuitive result simply follows from the symmetries of
the Hamiltonian expressed in the eigenbasis of coupled qubits, cfr 
eqs. \eref{H-trunc-SWAP}~-~\eref{H-inter}.  
As a consequence, the conditions 
to generate entangled states 
within the SWAP-subspace involve the spectra of transverse and longitudinal 
fluctuations at two different frequencies. 
An efficient $\sqrt{{\rm i-SWAP}}$ operation can be realized when
$S_{x \alpha} (\Omega) \ll S_{z \alpha}(\omega_c)$. 
The long-to-intermediate time behavior and the relevant decoherence times are 
summarized in \tref{table:decoherence}.
 
Defocusing processes instead originate both from transverse and longitudinal 
adiabatic noise. Remarkably, consideration of the coupled systems band structure
allows identification of operating conditions of reduced sensitivity to low frequency
fluctuations. An "optimal coupling" can be identified where defocusing is minimized. 
The error at the first $\sqrt{{\rm i-SWAP}}$ operation at optimal coupling can 
be reduced of a factor of $4$-to-$10$ with respect to operating at generic coupling strength.
The possibility of optimal tuning is ultimately due to the absence of effective longitudinal
noise in the SWAP-subspace and to the interplay of effective transverse intra- and inter-doublet
fluctuations. The orthogonal Z-subspace instead, because of the presence of effective
longitudinal noise, turns out to be much more sensitive to low-frequency noise which
cannot be limited by properly choosing system parameters. Therefore, populating this
subspace may severely limit the gate efficiency. 

The short-times behavior of the relevant dynamical quantities crucially depends
on adiabatic noise and its interplay with quantum noise. The dependence turns from quadratic
to linear depending on the largest component, parametrized by the SWAP-splitting
variance resulting from adiabatic noise, $\Sigma_{21}$, and the SWAP-decoherence rate
due to quantum noise $[2 T_1^{SWAP}]^{-1}$, as summarized in \tref{short}.

The considerable protection from 1/f noise achievable in the SWAP-subspace for optimal
coupling may suggest to use this subspace to encode a single quantum bit, similarly to
a decoherence-free-subspace~\cite{DFS}. Such an encoding would be convenient if the
SWAP decoherence time was about two orders of magnitude larger than the qubit decoherence 
time $T_2$. In fact,  because of the smaller SWAP oscillation frequency, 
$\omega_c \approx  10^{-1 (-2)}\Omega$,
the quality factor of a single qubit rotation within this subspace would
be  $Q= T_2^{SWAP} \omega_c \approx 10^{-1 (-2)} T_2^{SWAP} \Omega $, therefore 
$Q \gg T_2 \Omega $ if  $T_2^{SWAP} \gg 10^{1 (2)} T_2$.
Realizing this condition requires improving the relaxation times with respect to present day
experiments.  
A more realistic possibility is instead  to employ the SWAP subspace  
as a single qubit quantum memory. The requirement in this case would be
satisfying the weaker condition, $T_2^{SWAP} > T_2$.

Finally, we would like to comment on the effects of selected impurities strongly coupled to the
two-qubit gate. The detrimental effect of charged bistable fluctuators on single qubit 
charge or charge-phase gates has been observed in experiments~\cite{single-super,vion} 
and explained in theory~\cite{PRL05,Altshuler}. Effects on a two-qubit gate may even be
worse, as reported in the recent experiment on two coupled quantronium~\cite{nguyen}.
Explanation of the rich physics which comes out when selected impurities couple
to the nano-device is beyond the scope of the present article. In the following 
\ref{appendix:SC}, we illustrate the possible scenario when an impurity considerably changes the
4-level spectrum of the coupled qubits and we speculate on the possibility
to limit its effects by proper tuning the system parameters, somehow extending the optimal
tuning recipe. The usefulness of such a choice however critically depends
on the interplay with quantum noise at intermediate frequencies, an information 
unavailable in present day experiments. 
The analysis in \ref{appendix:SC} is the first step towards the identification of parameter 
regimes where a multilevel nano-device may be protected also from strongly coupled
degrees of freedom of a structured bath.

\ack
Stimulating discussions with G. Sch\"on and U. Weiss are gratefully acknowledged.

\appendix

\section{Capacitively coupled Cooper-Pair-Box based qubits}
\label{appendix-CPB}
\begin{figure}
\centering
\includegraphics[width=0.6\textwidth]{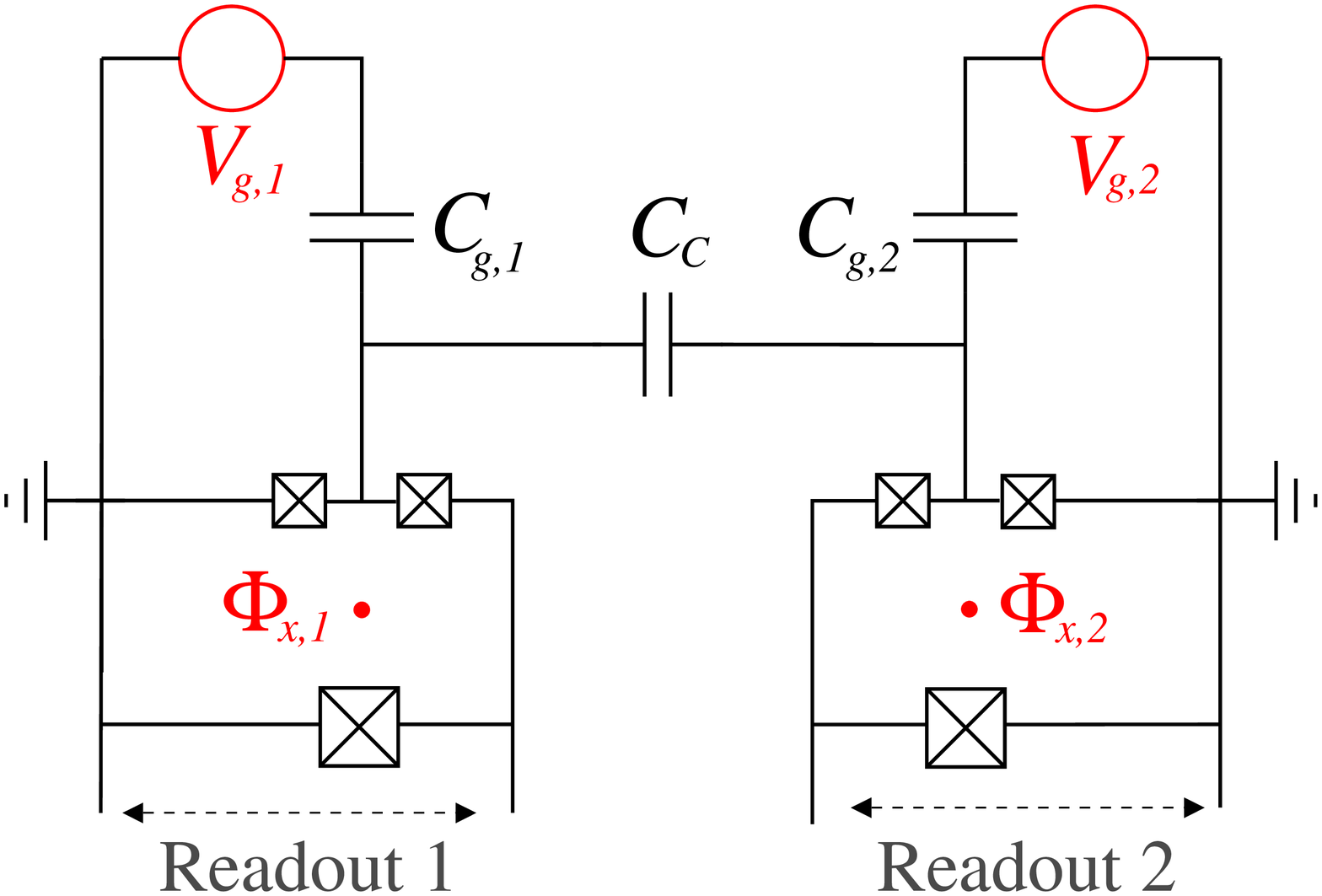}
\caption{Electrical circuit of two quantronia, labeled as 1 and
2, coupled via a fixed capacitor $C_\mathrm{C}$.
Each qubit has the characteristic two-port design, with control
parameters the gate voltage $V_{g, \alpha}$ and the magnetic flux
$\Phi_{x, \alpha}$ threading the superconducting loop~\cite{vion}.}
\label{fig:twocouqua}
\end{figure}

The Hamiltonian $\mathcal{H}_{0} $ in Eq. \eref{H0} 
is the central model of any non trivial two-qubit gate based on a fixed coupling
scheme. 
In particular, it describes coupled superconducting
qubits in the various implementations~\cite{coupled-exp-fix}. 
With $\mathcal{H}_{\rm I}$, given in \eref{eq:Hnoise}, it includes independent
fluctuations responsible for incoherent processes of different physical origin.
In this Section we derive $\mathcal{H}_{0} + \mathcal{H}_{\rm I}$
for capacitive coupled Cooper-Pair-Box-based (CPB) nano-devices. 
The CPB  is the main building block of many superconducting qubits~\cite{CPB}. 
Here we consider two charge-phase qubits (quantronia)~\cite{vion}
electrostatically coupled via a fixed capacitor, as schematically illustrated in \fref{fig:twocouqua}.

Each qubit is operated via two control parameters, the gate voltage $V_g$ and 
the magnetic flux across the junction loop, $\Phi_{x}$. They enter the CPB Hamiltonian via
the dimensionless parameters
$q_\mathrm{x}= C_\mathrm{g} V_g/(2e)$ and $\delta = \pi \Phi_x/\Phi_0$: 
\begin{equation}
	\mathcal{H}_{\mathrm{CPB}} = E_\mathrm{C}^0 (\hat{q}-q_\mathrm{x})^2 
	- E_\mathrm{J}(\delta) \cos \hat{\varphi},
\label{eq:single-ham}
\end{equation}
here  $E_\mathrm{C}^0 = 2e^2/C_\Sigma$ 
and $E_\mathrm{J}(\delta) = E_\mathrm{J}^0 \cos \delta$ is the Josephson 
energy, modulated by the phase $\delta$ around the zero phase value $E_\mathrm{J}^0$. 
Charge $\hat{q}$ and phase $\hat{\varphi}$ are conjugate operators, 
$[ \hat{\varphi}, \hat{q}]=\rmi$.
The loop capacitance $C_\Sigma$ is the sum of the gate $C_\mathrm{g}$ and the 
junctions $C_\mathrm{J}$ capacitances. 
The two CPBs are coupled by inserting a fixed capacitance 
$C_\mathrm{C}$ between the two islands. 
The presence of the coupling capacitance leads
to a renormalization of the CPB charging energies 
$E_{\alpha,\mathrm{C}} = E_{\alpha,\mathrm{C}}^0 ( 1 - C_\mathrm{T}/C_{\alpha,\Sigma} )$,
being
$E_{\alpha,\mathrm{C}}^0$ the CPB charging energy of the 
uncoupled qubit $\alpha$,
$1/C_\mathrm{T} = 1/C_\mathrm{C} + 1/C_{1,\Sigma} + 1/C_{2,\Sigma}$
the total inverse capacitance of the device. The coupling energy is 
$E_{\mathrm{CC}} = (2e)^2 C_\mathrm{T} / (C_{1,\Sigma} C_{2,\Sigma})$, 
and the full device Hamiltonian reads
\begin{eqnarray}
\mathcal{H}_\mathrm{D} = \mathcal{H}_{\mathrm{CPB} 1} \otimes \mathbb{I}_{2} + 
\mathbb{I}_{1} \otimes \mathcal{H}_{\mathrm{CPB} 2}  + \mathcal{H}_\mathrm{C} 
\label{eq:2qbham}
\\
\mathcal{H}_{\mathrm{CPB} \alpha}=
E_{\alpha,\mathrm{C}} 
[\hat{q}_\alpha - q_{\alpha,\mathrm{x}} \mathbb{I}_{\alpha}]^2 -
E_{\alpha,\mathrm{J}}(\delta_\alpha) \cos \hat{\varphi}_\alpha ,
\nonumber \\
\mathcal{H}_\mathrm{C}  = E_{\mathrm{CC}} [\hat{q}_1 - q_{1,\mathrm{x}} \mathbb{I}_{1} ] \otimes 
[\hat{q}_2 - q_{2,\mathrm{x}} \mathbb{I}_{2} ],
\label{eq:2qbham-coupl}
\end{eqnarray}
where the control parameters $q_{\alpha,\mathrm{x}}$ and $\delta_\alpha$, and
the charge $\hat{q}_\alpha$ and phase operators $\hat{\varphi}_\alpha$ play
the same role as in the single qubit case. 
At sufficiently low temperatures, the evolution of each CPB approximately takes place
within the bi-dimensional subspace spanned by the lowest energy eigenstates of 
$\mathcal{H}_{\mathrm{CPB} \alpha}$, $\vert \pm \rangle_\alpha$
with a splitting  depending on the control parameters \cite{varenna},
$\mathcal{H}_{\alpha} = \mathcal{P}_\alpha \mathcal{H}_{\mathrm{CPB} \alpha}
\mathcal{P}_\alpha = -\frac{1}{2} \Omega_\alpha \sigma_{\alpha z}$, where
the projection operator reads
$
\mathcal{P}_\alpha = \; _\alpha \! \vert + \rangle \langle + \vert_\alpha
+ \; _\alpha \! \vert - \rangle \langle -\vert_\alpha$ and
$\sigma_{\alpha z} = \;
_\alpha \! \vert - \rangle \langle -\vert_\alpha -
\;  _\alpha \! \vert + \rangle \langle +\vert_\alpha $.
The restricted dynamics of the coupled CPBs can be described in a pseudo-spin 
formalism by projection in the eigenstates basis  
$\{ | \mu,\nu \rangle = |\mu_1 \rangle \otimes | \nu_2 \rangle \}$. 
In this subspace the charge and Josephson operators are expressed in terms both 
of $\sigma_{\alpha z} $ and of the transverse component 
$\sigma_{\alpha x} = \; _\alpha \! \vert + \rangle \langle -\vert_\alpha
+ \; _\alpha \! \vert - \rangle \langle + \vert_\alpha$ as follows
\begin{eqnarray}
\!\!\!\!\!\!\!\!\! \!\!\!\!\!
\mathcal{P}_\alpha \hat{q}_\alpha \mathcal{P}_\alpha  = 
- \frac{1}{2} (q_{\alpha,++} - q_{\alpha,--}) \sigma_{\alpha z}
+ q_{\alpha,+-} \sigma_{\alpha x} + \frac{1}{2}
(q_{\alpha,++} + q_{\alpha,--}) \mathbb{I}_{\alpha}
\label{eq:pcharge} \\
\!\!\!\!\!\!\!\!\!\! \!\!\!\!\!\!\!\!\!\!\!\!\!\!\!\!\!\!
\mathcal{P}_\alpha (\cos \hat{\varphi}_\alpha) \mathcal{P}_\alpha  = 
- \frac{1}{2} (\Phi_{\alpha,++} - \Phi_{\alpha,--}) \sigma_{\alpha z}
+ \Phi_{\alpha,+-} \sigma_{\alpha x}+
\frac{1}{2} (\Phi_{\alpha,++} + \Phi_{\alpha,--})\mathbb{I}_{\alpha},
\label{eq:pphase}
\end{eqnarray}
where 
\begin{equation*}
q_{\alpha, \mu\nu} = q_{\alpha, \nu\mu} = \;
_\alpha \!  \langle \mu \vert \hat{q}_\alpha \vert \nu \rangle_\alpha 
\qquad  \qquad \Phi_{\alpha, \mu\nu} = \Phi_{\alpha, \nu\mu} = \;
_\alpha \!  \langle \mu \vert \cos \hat{\varphi}_\alpha \vert \nu \rangle_\alpha \, .
\end{equation*}
The restriction of the device Hamiltonian to $\{ | \mu,\nu \rangle\}$ 
is \footnote{We have used the shorthand notation $\sigma_\alpha$  to indicate 
$\sigma_\alpha \otimes \mathbb{I}_{\beta}$, 
$\alpha \neq \beta$.}
\begin{eqnarray}
(\mathcal{P}_1 \otimes \mathcal{P}_2) \mathcal{H}_\mathrm{D}
	       (\mathcal{P}_1 \otimes \mathcal{P}_2) \,  \equiv \, \tilde \mathcal{H}_\mathrm{D} =
	        - \sum_\alpha \frac{\Omega_\alpha }{2} \sigma_{\alpha z}  +
		 \nonumber \\
\!\!\!\!\! \!\!\!\!\!  + \frac{E_{\mathrm{CC}}}{2} 
	       \sum_{\alpha \neq \beta} \left [- \frac{1}{2} (q_{\alpha,++} - q_{\alpha,--}) \sigma_{\alpha z}
	      + q_{\alpha,+-} \sigma_{\alpha x} \right ] (q_{\beta,++} + q_{\beta,--}- 2 q_{\beta,x}) 
	      \nonumber \\
\!\!\!\!\! \!\!\!\!\! 
      + E_{\mathrm{CC}} \prod_\alpha  \left [ - \frac{1}{2} (q_{\alpha,++} - q_{\alpha,--}) \sigma_{\alpha z}
	      + q_{\alpha,+-} \sigma_{\alpha x} \right]
\label{eq:htrunc}
\end{eqnarray}
Note that because of the coupling, in addition to interaction terms between the two qubits, each
qubit is effectively displaced from its own operating point (second term in \eref{eq:htrunc}).

Because of fluctuations of control parameters the device couples to the external
environment. The form of interaction terms can be easily deduced considering classical
fluctuations of  $q_{\alpha,\mathrm{x}}$ and $\delta_\alpha$, i.e. by replacing in \eref{eq:2qbham} and 
\eref{eq:2qbham-coupl},
$q_{\alpha,\mathrm{x}} \to q_{\alpha,\mathrm{x}} + \Delta q_{\alpha,\mathrm{x}}$ and 
$\delta_\alpha \to \delta_\alpha + \Delta \delta_\alpha $.
We therefore obtain $\mathcal{H}_\mathrm{D} \to \mathcal{H}_\mathrm{D} + 
\delta \mathcal{H}_\mathrm{D}$, where
\begin{equation*}
\delta \mathcal{H}_\mathrm{D} = \delta \mathcal{H}_{\mathrm{CPB}1} \otimes
\mathbb{I}_{2} + \mathbb{I}_{1} \otimes \delta \mathcal{H}_{\mathrm{CPB} 2}
+\delta \mathcal{H}_\mathrm{C}.
\end{equation*}
Fluctuations of each CPB Hamiltonian and of the coupling term read
\begin{eqnarray}
\delta \mathcal{H}_{\mathrm{CPB} \alpha} = X_\alpha \hat{q}_\alpha + 
Z_\alpha \cos \hat \varphi_\alpha 
\label{eq:flucthdev}\\ 
\delta \mathcal{H}_\mathrm{C}= g_2 X_2 \, \hat{q}_1 \otimes \mathbb{I}_{2} + g_1 X_1 
\mathbb{I}_{1} \otimes \hat{q}_2 \, ,
\end{eqnarray}
where $X_\alpha$ is related to gate charge fluctuations,
$X_\alpha = - 2 E_{\alpha,\mathrm{C}} \Delta q_{\alpha,\mathrm{x}}$ 
and $Z_\alpha$ to fluctuations of the  phase or equivalently of the
Josephson energy, $Z_\alpha = - \Delta E_{\alpha,\mathrm{J}}$, where
$\Delta E_{\alpha,\mathrm{J}} = E_{\alpha,\mathrm{J}}(\delta_\alpha+ \Delta \delta_\alpha)
 -  E_{\alpha,\mathrm{J}}(\delta_\alpha) 
$, and we put  $g_\alpha = E_{CC}/(2 E_{\alpha,C})$.
In the computational subspace the additional terms reduce to 
\begin{eqnarray}
\!\!\!\!\!\delta \mathcal{H}_{\alpha} = 
- \frac{1}{2} X_\alpha (q_{\alpha,++} - q_{\alpha,--}) \sigma_{\alpha z}
+ X_\alpha q_{\alpha,+-} \sigma_{\alpha x} + \frac{1}{2} X_\alpha
(q_{\alpha,++} + q_{\alpha,--}) \mathbb{I}_{\alpha}  \nonumber \\
\!\!\!\!\!- \frac{1}{2} Z_\alpha (\Phi_{\alpha,++} - \Phi_{\alpha,--}) \sigma_{\alpha z}
+ Z_\alpha \Phi_{\alpha,+-} \sigma_{\alpha x} +
\frac{1}{2} Z_\alpha (\Phi_{\alpha,++} + \Phi_{\alpha,--})\mathbb{I}_{\alpha}
\end{eqnarray}
and 
\begin{eqnarray}
\!\!\!\!\!\delta \mathcal{H}_\mathrm{C} = 
g_2 X_2 \left [ - \frac{1}{2} (q_{1,++} - q_{1,--}) \sigma_{1 z}
+ q_{1,+-} \sigma_{1 x} + \frac{1}{2}
(q_{1,++} + q_{1,--}) \mathbb{I}_{1} \right ] \otimes \mathbb{I}_{2} 
\nonumber \\
\!\!\!\!\!+ g_1 X_1 \mathbb{I}_{1} \otimes \left [ - \frac{1}{2} (q_{2,++} - q_{2,--}) \sigma_{2 z}
+ q_{2,+-} \sigma_{2 x}+ \frac{1}{2}
(q_{2,++} + q_{2,--}) \mathbb{I}_{2} \right ]  .
\end{eqnarray}
In conclusion, the coupled CPBs Hamiltonian at a general working point, in the 
four dimensional  subspace $\{ | \mu,\nu \rangle\}$,  
including fluctuations of the control parameters and neglecting constant terms, takes the
form $\tilde\mathcal{H}_\mathrm{D} +  \delta \tilde \mathcal{H}_\mathrm{D}$ with 
$\tilde\mathcal{H}_\mathrm{D} $ given in eq. \eref{eq:htrunc} and
\begin{eqnarray}
\!\!\!\!\! \!\!\!\!\! \!\!\!\!\! \!\!\!\!\! \!\!\!\!\! \!\!\!\!\! \!\!\!\!\!\!\!\!\!\!
\delta \tilde\mathcal{H}_\mathrm{D} =
\sum_\alpha \left [\frac{1}{2} \left ( X_\alpha (q_{\alpha,--} - q_{\alpha,++})
+ Z_\alpha (\Phi_{\alpha,--} - \Phi_{\alpha,++}) \right ) \sigma_{\alpha z}
+ (X_\alpha q_{\alpha,+-} + Z_\alpha \Phi_{\alpha,+-})\sigma_{\alpha x}  \right]
\nonumber \\
\!\!\!\!\! \!\!\!\!\! \!\!\!\!\! \!\!\!\!\! \!\!\!\!\! \!\!\!\!\! \!\!\!\!\!
+ g_2 X_2 \left [\frac{1}{2} (q_{1,--} - q_{1,++}) \sigma_{1 z}
+ q_{1,+-} \sigma_{1 x} \right ]
+ g_1 X_1 
\left [ \frac{1}{2} (q_{2,--} - q_{2,++}) \sigma_{2 z}
+ q_{2,+-} \sigma_{2 x} \right ]  
\end{eqnarray}
Note that due to the capacitive coupling a cross-talk effect takes place, gate charge fluctuations
of qubit $\alpha$ being responsible for fluctuations of the polarization of qubit $\beta$. Effects
are scaled with the coupling energy $E_{\mathrm{CC}}$,
thus they are expected to be less relevant compared to fluctuations acting directly on each qubit. 
A detailed analysis of cross-talk and correlations
between gate charge fluctuations has been reported in \cite{NJP-special}. 
In our analysis we disregarded cross-talk and correlations between noise sources acting on each qubit.

\begin{table}
\caption{\label{tab:numbers} Typical values of the CPBs parameters \cite{nguyen}. 
$\Omega^0$ is the energy splitting of the single qubit.}
\begin{indented}
\item[]
\begin{tabular}{lll}
\br
Parameter & Qubit 1 & Qubit 2  \\
\mr
$E_\mathrm{C}^0$ (GHz) & 9.997 & 10.41 \\
$E_\mathrm{J}^0$ (GHz) & 14.16 & 15.62 \\
$C_\Sigma$ (fF) & 7.73 & 7.42 \\
$\Omega^0$ (GHz) & 12.666 & 13.81 \\
$E_\mathrm{\alpha,C}$ (GHz) & 9.92 & 10.33 \\
$\Omega$ (GHz) & 12.645 & 13.79 \\
\br
\end{tabular}
\item[]
\end{indented}
\end{table}
\subsection{Choice of the working point}
The implementation of a two qubit gate in a fixed coupling scheme
requires the qubits to be at resonance during gate operation.
Because of experimental tolerances on bare parameters (about $10$\%),
the resonant condition can only be achieved by a proper choice of the
single-qubit working points via tuning $q_{\alpha, \mathrm{x}}$ and $\delta_\alpha$.
This is a critical choice since at least one qubit has to be moved away from the working 
point of minimal sensitivity to parameters variations, the "optimal point", 
$q_{\alpha, \mathrm{x}}=1/2$, $\delta_\alpha=0$~\cite{vion}. 
Since CPB-based devices are severely affected by low-frequency charge noise 
it is desirable to maintain the charge optimal point, $q_{\alpha,\mathrm{x}}=1/2$
and modulate resonance/detuning  by tuning the phases $\delta_\alpha$.
The most reasonable choice consists in maintaining one qubit at its own phase optimal point,
$\delta_1=0$ and  tune the phase of the other qubit,  $\delta_2$.
For set of parameters close to those planned in experiments~\cite{nguyen} 
(see \tref{tab:numbers})
resonance occurs for $\delta_2 \approx 0.455$.   
Since at $q_{\alpha,\mathrm{x}}=1/2$ it results 
$q_{\alpha,++} = q_{\alpha,--} = 1/2$ and $\Phi_{\alpha,+-}=0$, 
 the projected charge \eref{eq:pcharge} is transverse
and the Josephson operator \eref{eq:pphase}
is longitudinal with respect to each Hamiltonian $\mathcal{H}_{\alpha}$.
Therefore, the truncated Hamiltonian is of the form $\mathcal{H}_0 + \mathcal{H}_{\mathrm{I}}$
given by eqs. \eref{H0} and \eref{eq:Hnoise} respectively
\begin{eqnarray}
\fl \mathcal{H}_0 = 
 -\frac{\Omega}{2} \, \sigma_{1z } \otimes \mathbb{I}_{2}
 -\frac{\Omega}{2}  \, \mathbb{I}_{1} \otimes \sigma_{2 z}
+E_\mathrm{CC} \, \tilde q_{1,+-} \, \tilde q_{2,+-} \, \sigma_{1 x} \otimes \sigma_{2x} 
\label{eq:hresonance}\\
\fl \delta \mathcal{H}_0 =
\sum_{\alpha \neq \beta} \left [
\frac{1}{2} Z_\alpha (\tilde \Phi_{\alpha,--} - \tilde \Phi_{\alpha,++})  \sigma_{\alpha z}
+ X_\alpha \tilde q_{\alpha,+-} \sigma_{\alpha x}  \right] \otimes \mathbb{I}_{\beta}
\label{eq:hreson-noise} 
\end{eqnarray}
where $\tilde q_{\alpha,+-}$, $\tilde \Phi_{\alpha,++}$ etc. denote the specific values 
taken by these matrix elements at the resonance point obtained for a specific choice of $\delta_2$.
Values of the charge and phase matrix elements at this working point are reported in
table \ref{tab:meanvalues} and lead to $E_\mathrm{CC}=0.18$ GHz.
The interaction Hamiltonian $\mathcal{H}_{\mathrm{I}}$ results from $\delta \mathcal{H}_0$
considering the quantized version of the classical fluctuations 
$x_\alpha = -4 E_{\alpha,\mathrm{C}} \tilde q_{\alpha,+-} \Delta  q_{\alpha,\mathrm{x}}$
and $z_\alpha =  (\tilde \Phi_{\alpha,++} - \tilde \Phi_{\alpha,--}) \Delta E_{\alpha,\mathrm{J}}$.
Where, since $\delta_1=0$ we have $\Delta E_{1, \mathrm{J}} = - 0.5 E_{1,\mathrm{J}}^0 (\Delta  \delta_1)^2$ 
whereas for $\delta_2 \neq 0$ it results 
$\Delta E_{2, \mathrm{J}} = -  E_{2,\mathrm{J}}^0 (\sin \delta_2) \, \Delta  \delta_2$.
\begin{table}
\caption{\label{tab:meanvalues} Matrix elements of the charge and Josephson
operators in the two lowest eigenstates basis of each CPB Hamiltonian.
The physical parameters of each CPB are given in \tref{tab:numbers}.}
\begin{indented}
\item[]\begin{tabular}{ccc}
\br
 & Qubit 1 & Qubit 2  \\
\mr
$\tilde q_{\alpha, ++}- \tilde q_{\alpha, --}$ & 0 & 0 \\
$\tilde q_{\alpha, +-}$ & -0.5962161 & -0.5891346 \\
$\tilde \Phi_{\alpha, ++}- \tilde \Phi_{\alpha, --}$ & 0.7049074 & 0.7396952 \\
$\tilde \Phi_{\alpha, +-}$ & 0 & 0 \\
\br
\end{tabular}
\end{indented}
\end{table}

\begin{figure}
\centering
\resizebox{0.5\columnwidth}{!}{
\includegraphics{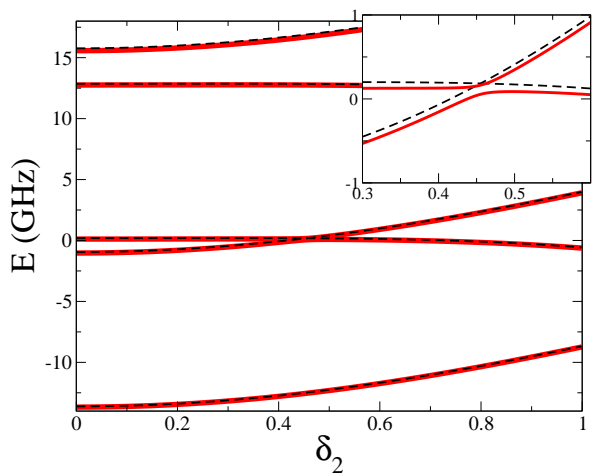}} 
\caption{The five lowest eigenenergies of the coupled two CPB
Hamiltonian (\ref{eq:2qbham}) at charge optimal points 
$q_{1,\mathrm{x}}= q_{2,\mathrm{x}}=1/2$
versus the qubit 2 phase $\delta_2$, being $\delta_1=0$. The dashed black and red lines 
refer to the uncoupled and coupled case, respectively.  
The $5$-th eigenvalue lies sufficiently far in energy to be neglected in our analysis.
In the inset the 
degeneracy point is shown with the small anti-crossing due to the coupling term
at $\delta_2 = 0.455$.}
\label{fig:two-split}
\end{figure}

The accuracy of the truncation of the multistate device Hamiltonian to the lowest
four eigenstates has been checked numerically by exact diagonalization of 
\eref{eq:2qbham} using the numerical values of the physical  parameters reported 
in \tref{tab:numbers} and considering 14 charge states for each qubit, see \fref{fig:two-split}. 

\subsection{Structured noise in CPB-based circuits}
\label{subsec:noise}

Fluctuations of the control parameters $q_{\alpha,\rm{x}}$, $\delta_\alpha$
have different physical origin. 
They partly stem from microscopic noise sources, partly from the circuitry
itself.  The resulting stochastic processes are sometimes non-Gaussian, therefore 
complete characterization of the processes requires in principle higher-order cumulants. 
Such a complete description is often unavailable in 
experiments, whereas it is usually possible to provide a characterization of
the power spectrum of the various stochastic processes. 
Noise characteristics for the two quantronia are summarized in table \ref{tab:noise}
(\sref{sec:quantumnoise}).

Experiments on various Josephson implementations have revealed the presence of spurious 
resonances in the spectrum which manifest themselves as beatings in time resolved 
measurements~\cite{vion}.
In charge and charge-phase qubits they are due to strongly coupled (SC) background charges and
may severely limit the reliability of these devices~\cite{PRL02,Altshuler}. In the following
\ref{appendix:SC}
we will discuss the detrimental effect of a single strongly-coupled
impurity on the visibility of a $\sqrt{{\rm i-SWAP}}$ operation.

Finally, we mention that in the present analysis we have not explicitly considered 
fluctuations of the Josephson energy due to critical current fluctuations. 
The effect of $E_\mathrm{J}^0$  noise %low-frequency Josephson energy noise 
may be straightforwardly included in our analysis within the multistage approach.

\section{Effect of a strongly coupled impurity}
\label{appendix:SC}

Experiments on various Josephson implementations have revealed the presence of spurious 
resonances in the spectrum which manifest themselves as beatings in time resolved 
measurements~\cite{vion,Simmonds,Ustinov,resonators}.
In charge and charge-phase qubits they are due to strongly coupled (SC) background charges and
may severely limit the reliability of these devices~\cite{PRL02,Altshuler}.
Several experiments have shown that these impurities can be modeled as bistable fluctuators (BF),
switching between states $0$ and $1$ with a rate $\gamma$. Often the qubit-BF coupling is
transverse, i.e. of the form $- (v/2) \sigma_x \xi(t)$ with $\xi(t)=\{ 0,1 \}$~\cite{ustinovAPL}. Each bistable state
corresponds to a different qubit splitting, $\Omega$ for $\xi=0$ and 
$\Omega^\prime \approx \Omega[1+ 0.5 (v/\Omega)^2] $ for $\xi=1$ ($v \ll \Omega$). 
The parameter quantifying the strength of the qubit-BF coupling  is 
$g_1 \equiv (\Omega^\prime - \Omega)/\gamma$~\cite{PRL02}. 
When $g_1 \gg 1$ the impurity is strongly coupled and it is visible both in 
spectroscopy and in the time-resolved dynamics.
Similarly, a single BF acting on one of the two qubits forming a universal gate 
induces a bi-stability in the SWAP-splitting: $\omega_{21} = \omega_c$ for $\xi=0$ and 
$\omega_{21}^\prime \approx \omega_c[1- 0.5 (v/\Omega)^2  + v^4/ (8\omega_c^2 \Omega^2)]$ 
for $\xi=1$ (in this case the stronger condition on the qubit-BF coupling has
been assumed $v \ll \omega_c$).
The strong coupling condition for the $\sqrt{{\rm i-SWAP}}$ gate is therefore
$g_2 = (\omega_{21}^\prime  - \omega_{21})/\gamma \gg 1$. Considering the above expansions 
and the constraint $\Omega > \omega_c$ the two ratios satisfy the condition
$g_1 > g_2$. Therefore the visibility of a specific BF is expected to be reduced in a two-qubit
operation with respect to a single qubit gate. 

\begin{figure}
\begin{center}
\includegraphics[width=0.47\textwidth]{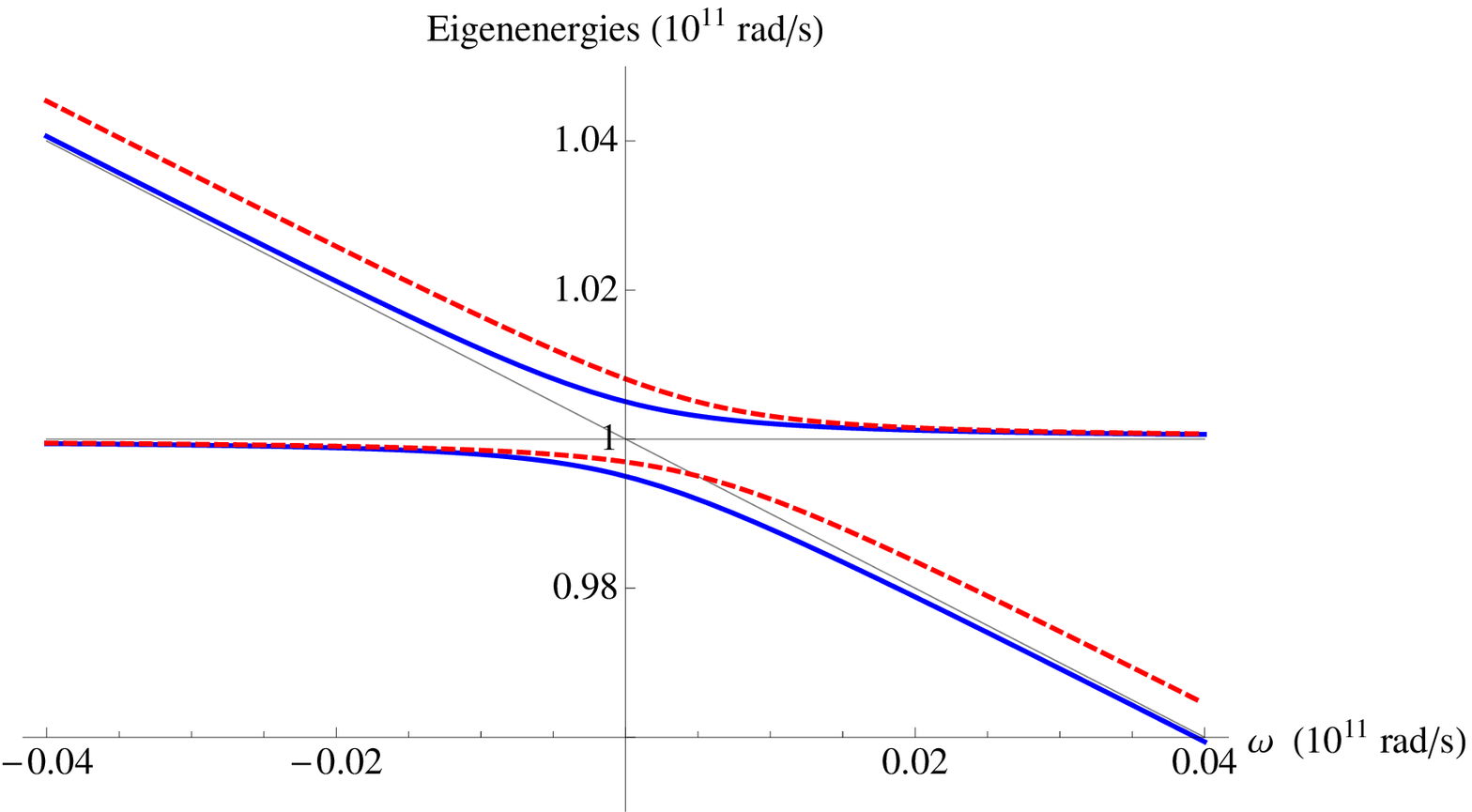}
\includegraphics[width=0.45\textwidth]{figureB1b}
\end{center}
\caption{
Left: Simulated spectroscopy of the SWAP splitting $\omega_{21}$  for  qubit-qubit 
coupling $\omega_c/\Omega=0.01$ (blue) and for uncoupled qubits (black). 
In the presence of a  BF  coupled to one qubit with strength $v/\Omega=0.1$
the coupled qubits ($\omega_c/\Omega=0.01$) eigenenergies are shifted (red).
Right: Numerical simulations of the switching probabilities in the presence 
of the above BF (orange/cyan), and of the BF plus $1/f$ noise with 
$\sigma_x=0.02 \Omega$ (blue and red). 
For optimal coupling, $\tilde \omega_c= \sqrt{\sigma_x^2 + v^2/4}$, a considerable 
recover of the signal can be obtained (gray).
}\label{fig-BF}
\end{figure}

The effect of a single BF fluctuator depends on the way experimental data are collected.
For spectroscopic measurements the single-shot time is about $200$~ns and
the acquisition time of a single point in the spectrum requires about $4 \times 10^4$
repetitions, resulting in a recording time for a single data point 
of about $10^{-2}$~s \cite{nguyen}. Therefore
a single BF will be visible in spectroscopy if its switching time is
$0.5 \times 10^{-7} {\rm s} < 1/\gamma < 2 \times  10^{-2} {\rm s} $. For instance a BF 
switching at $\approx 10$~kHz can be averaged during the spectroscopic measurements and
it is visible if it is sufficiently strongly coupled to one qubit, for instance this is
the case for  $v/\Omega =0.1 > \omega_c/\Omega $. The expected effect of this 
impurity in spectroscopy is shown in \fref{fig-BF} where it results in the simultaneous 
presence of two avoided crossings depending on the impurity state.  
The occurrence of a similar BF represents a major problem for charge and charge-phase 
implementations~\cite{nguyen}. Effects are also visible in time resolved measurements.
If the considered BF is one out of the several ones responsible for $1/f$ noise, the global effect
of the structured bath is a considerable reduction of the oscillation amplitude, as shown in
\fref{fig-BF}.
Under these conditions, in principle, an "optimal coupling" can still be identified. Since the 
main problem is the beating pattern, optimal tuning is defined by the condition that
the average $\langle \omega_{21}^\prime  - \omega_{21} \rangle$
vanishes, rather than minimizing the SWAP-splitting variance. 
In the presence of only charge noise, this condition leads to a modified "optimal coupling", 
$\tilde \omega_c= \sqrt{\sigma_x^2 + v^2/4}$, at 
which effectively the SWAP visibility is improved (\fref{fig-BF}). 
The efficiency of such a choice however critically depends on the details (presently not available)
of transverse  noise at frequencies in the kHz~-~MHz range. 
Relaxation processes due to quantum noise at these frequencies may in fact represent an
additional liming factor for the gate fidelity. 

\section{Frequency shifts}
\label{appendix:shifts}

In this Appendix we report the frequency shifts entering the off-diagonal
elements of the RDM as resulting from the solution of the Master Equation
in the secular approximation reported in \sref{sec:quantumnoise}.
The frequency shifts take the simple form
\begin{eqnarray}
\begin{array}{ll}
\tilde\omega_{12} - \omega_{12} \, = \, \frac{1}{2} \,
\left [
\sum_{k \neq 2} {\mathcal E}_{2k} - \sum_{k \neq 1} {\mathcal E}_{1k} 
\right ] \\
\tilde \omega_{03} - \omega_{03} \,= \,\frac{1}{2} \,
\left [\sum_{k \neq 0} {\mathcal E}_{0k} 
- \sum_{k \neq 3} {\mathcal E}_{3k} \right] \,.
\end{array}
\label{shifts}
\end{eqnarray}
For the SWAP coherence the different contributions read
\begin{eqnarray}
{\mathcal E}_{10} &=&  \frac{1}{8} \, (1+\sin \varphi) \,[{\mathcal E}_{x_1}(\omega_{10}) +
{\mathcal E}_{x_2}(\omega_{10})] \nonumber \\
{\mathcal E}_{13} &=& \frac{1}{8} \, (1-\sin \varphi) \,[{\mathcal E}_{x_1}(\omega_{13}) +
{\mathcal E}_{x_2}(\omega_{13})] 
\nonumber \\
{\mathcal E}_{20} &=& \frac{1}{8} \, (1-\sin \varphi) \,[{\mathcal E}_{x_1}(\omega_{20}) +
{\mathcal E}_{x_2}(\omega_{20})] 
\nonumber \\
{\mathcal E}_{23} &=& \frac{1}{8} \, (1+\sin \varphi) \,[{\mathcal E}_{x_1}(\omega_{23}) +
{\mathcal E}_{x_2}(\omega_{23})] 
\nonumber \\ 
{\mathcal E}_{21} &=& \frac{1}{8} \{ [ \mathcal{E}_{z_1} (\omega_{21})
 + \mathcal{E}_{z_2} (\omega_{21})]
\nonumber \\
 {\mathcal E}_{12} &=&\frac{1}{8} \{ [ \mathcal{E}_{z_1} (\omega_{12}) 
+  \mathcal{E}_{z_2} (\omega_{12})]
 \nonumber 
\end{eqnarray}

From single qubit measurements, as reported in \sref{subsec:noise}, charge noise
at frequencies of order of $\Omega$ is white. 
Assuming that also the phase variables $\Delta \delta_\alpha$  have white spectrum at 
frequencies of order $\omega_c$, results in 
$S_{z_2}(\omega) =  ( E_{J,2}^0 \sin \delta_2 )^2 S_{\delta_2}(\omega) 
\approx 5 \times 10^7$s$^{-1}$. 
The spectrum of $\hat z_1$  takes instead  
the ohmic form $S_{z_1}(\omega) = 2 S^2/(\pi \Omega^2) \, \omega \, 
\coth(\omega / (2 K_B T))$, where $S= 1.6\times 10^8$s$^{-1}$.
Under these conditions, phase shifts due to polarization noise and phase fluctuations
on qubit 2 identically vanish (from \eref{principalv}). A frequency shift contribution results from ohmic phase noise
on qubit 1, similarly to a qubit affected by transverse noise~\cite{weiss}, as it can 
be argued from \eref{H-trunc-SWAP}
\begin{equation}
\!\!\!\!\!\!\!\!\!\!\!\!\!\!\!\!\!\!\!\!\!\tilde\omega_{12} - \omega_{12} \, =
\frac{1}{8}  \, [ \mathcal{E}_{z_1} (\omega_{21})
 - \mathcal{E}_{z_1} (- \omega_{21})] = {\mathcal P}
 \int_{-\infty}^{\infty} \frac{d \omega}{4\pi} \, \frac{S_{z_1}(\omega)}{1+e^{-\omega/k_B T}} \,
\frac{\omega_{21}}{\omega^2 -\omega_{21}^2} \,.
\label{shift-phase1}
\end{equation}
The shift depends on the high-frequency cut-off of the spectrum and
on the temperature. Explicit forms are known in the literature, see ~\cite{weiss}.
The damping strength parameter entering the ohmic spectrum, commonly denoted as $K$,
is in the present case $K= S^2/(\pi \Omega^2) \approx 10^{-6}$. The SWAP-splitting shift
scales with $K$ and it  will therefore be extremely small.
Possible logarithmic divergences with the high-frequency cut-off are not included in this
analysis since there's presently no indication on the features of the spectrum at those
frequencies.
For this reason frequency shifts  have been neglected both in Section
\ref{sec:quantumnoise} and in Section \ref{sec:adiab}.

\end{document}